\documentclass[aps, twocolumn]{revtex4-1}
\usepackage{graphicx}
\usepackage{dcolumn}
\usepackage[utf8]{inputenc}
\usepackage[T1]{fontenc}
\usepackage{bm}
\usepackage{ulem}
\usepackage{mathptmx}
\usepackage{amssymb}
\usepackage{etoolbox}
\usepackage{amsmath, amsfonts}
\usepackage{tikz}
\usepackage{enumerate}
\usepackage{url}
\usepackage{float}
\usepackage{wasysym}
\usepackage[mathscr]{eucal}
\usepackage{hyperref}
\usepackage{svg}
\usepackage{wrapfig}
\hypersetup{
    colorlinks=true,
    linkcolor=blue,
    filecolor=magenta,      
    urlcolor=cyan,
    }

\usetikzlibrary{shapes.geometric}
\usetikzlibrary{decorations.pathreplacing, angles, quotes}
\usetikzlibrary{arrows, decorations.markings}
\usetikzlibrary{plotmarks}
\usepackage{rotating,array}
\usepackage{xcolor}
\usepackage[utf8]{inputenc}
\definecolor{MATLABblue}{HTML}{0072BD}
\definecolor{MATLABorange}{HTML}{EDB120}
\definecolor{CTD}{rgb}{0.25,0.65,0.25}
\newtheorem{definition}{Definition}

\begin{document}
\title{Generalizing Geometric Partition Entropy for the Estimation of \\Mutual Information in the Presence of Informative Outliers}

\author{C. Tyler Diggans}
\affiliation{Air Force Research Laboratory's Information Directorate, 525 Brooks Road, Rome, NY 13441}

\author{Abd AlRahman R. AlMomani} 
\affiliation{Embry-Riddle Aeronautical University,
3700 Willow Creek Rd, Prescott, AZ 86301}

\begin{abstract}
    The recent introduction of geometric partition entropy brought a new viewpoint to non-parametric entropy quantification that incorporated the impacts of informative outliers, but its original formulation was limited to the context of a one-dimensional state space.  A generalized definition of geometric partition entropy is now provided for samples within a bounded (finite measure) region of a $d$-dimensional vector space. The basic definition invokes the concept of a Voronoi diagram, but the computational complexity and reliability of Voronoi diagrams in high dimension make estimation by direct theoretical computation unreasonable.  This leads to the development of approximation schemes that enable estimation that is faster than current methods by orders of magnitude.  The partition intersection ($\pi$) approximation, in particular, enables direct estimates of marginal entropy in any context resulting in an efficient and versatile mutual information estimator. This new measure-based paradigm for data driven information theory allows flexibility in the incorporation of geometry to vary the representation of outlier impact, which leads to a significant broadening in the applicability of established entropy-based concepts.  The incorporation of informative outliers is illustrated through analysis of transient dynamics in the synchronization of coupled chaotic dynamical systems.
\end{abstract}

\maketitle

\textbf{Estimating the entropy of an unknown distribution from a sample of observations in a bounded continuous state space has several well-known challenges, which only become more problematic as the dimensionality of the space increases. A new approach to the estimation of fundamental information measures, such as the mutual information, is provided. This makes use of the concept of partition (i.e. metric) entropy, which is common in symbolic dynamics, over data-driven partitions of the state space. Those partitions are best defined through coarse-graining the Voronoi diagram, but more efficient estimators are realized through several rectangular partitioning procedures.  The intentional choice of the measure used, for instance focusing on the geometry, can enable time series analysis that emphasizes the impact of outliers instead of averaging them away as other methods do, which enables new forms of perturbation analysis.  Of particular interest are the insights gained into the transient dynamics encountered in the study of the process of synchronization.  Other choices of measure can still provide estimates that correspond to traditional methods as well, but those are obtained at a fraction of the computational costs of common methods like the $k$-nearest neighbor estimator of mutual information.}
\section{Introduction}
Entropy-based analysis of continuous data, most commonly understood in the form of Differential Shannon Entropy (DSE), has been a cornerstone of information theory since its introduction by Claude Shannon in the last chapter of his seminal work~\cite{shannon1948mathematical}. However, DSE often provides a poor quantification of uncertainty \cite{paninski2003estimation,ciprian2012on,silva2012convergence,ricci2021asymptotic} from data for unknown probability distributions when i) the sample is sparse and/or ii) significant outliers are present.  Despite these shortcomings, if the sample size is orders of magnitude larger than the number of bins used in a histogram approximation of the probability density function (\textit{pdf}), such an approach can still provide useful estimates.  However, in higher dimensional settings, including the estimation of mutual information and many machine learning applications, the sample sizes required rapidly outpace most reasonable computational resources due to the ``curse of dimensionality'' associated with a changing concentration of measure~\cite{sricharan2013ensemble, goldfeld2018estimating}.  The successful application of these concepts was therefore restricted in many ways for several decades, especially in high dimensional settings, by either a need for \textit{a priori} knowledge of the underlying distribution for parametric estimation or large amounts of data to enable representative histogram approximations of a continuous \textit{pdf}. 

Significant advancements have been made possible in more recent decades with the introduction of the so-called geometric entropy estimators, which are grounded in $k$-nearest neighbor (knn) frequency analysis and epitomized by the renowned Kraskov–St\"{o}gbauer–Grassberger (KSG) estimator for mutual information \cite{ksg2004estimating}.  This statistical approach provided a more reliable non-parametric estimator for the higher dimensional measures of conditional mutual information and transfer entropy, which particularly impacted the analysis of data from coupled dynamical systems.  

The promise of these advances has been tempered, however, by their considerable computational complexity and the limitations of statistical analysis that has restricted their successful application to only those systems with a relatively small number of nodes and interactions~\cite{zhao2019analysis}.  Moreover, when mutual information is being estimated between strongly dependent variables, a notable issue with the knn estimators, in particular, is their requirement for prohibitively large sample sizes for accurate estimation~\cite{gao2015efficient}.  In cases where data is sparse or unevenly distributed across different dimensions, this limitation can lead to inaccuracies.

Finally, a thorough examination of the popular KSG method has indicated that its purported ``robustness to outliers'' may in fact be problematic in many applications. The general perception of robustness in this context is often conflated with the method's reliance on the law of large numbers, which suggests robustness is achieved not through an inherent resistance to outliers but rather through a statistical averaging process that diminishes the impact of these deviations. Such a characteristic, while seemingly advantageous, betrays unstated assumptions about the smoothness of the underlying distribution or the uniformity of information provided by each observation.  In practical scenarios, such assumptions may be warranted, but the presence of outliers is not necessarily a statistical anomaly; in fact, their presence can be indicative of low probability but critical aspects of the underlying phenomena recorded in the dataset, especially when analyzing dynamic data. Disregarding these outliers entirely, especially without a scientifically validated rationale, risks a loss of information that could be pivotal in understanding the behavior of many complex systems. This is especially pertinent in cases with fast and slow dynamics, switching, and/or chaotic systems.

As such, the idea, which is inherent to all frequency-based analysis including the KSG method, that the impact of outliers can be minimized based on large sample sizes is generally flawed.  In addition, obtaining large datasets in many real-world applications is either impractical or impossible, thereby compromising the reliability and applicability of the KSG estimator in such scenarios. Consequently, this method and its adaptations have limited efficacy for telling the full story about the dynamics that created certain data sets, particularly when dealing with time series from complex systems where every data point, including outliers, could provide meaningful insights.  Thus, the KSG method's approach to handling outliers, though often touted as a strength, reveals itself to be a fundamental liability for many applications.

A new approach to entropy estimation from a sample in the context of a one dimensional continuous state space was introduced by the authors in~\cite{diggans2022geometric}, and named Geometric Partition Entropy (GPE).  That approach was specifically designed to overcome the known limitations of traditional entropy estimators with respect to the handling of outliers, limited sample sizes, and/or atypical distributions.  A set of coarse-grained, evenly represented, but unequal sized macrostates that partition the state space were obtained by utilizing the quantiles of the sample.  The original GPE estimate was then defined as what can be described as the partition (or metric) entropy of this quantile-defined partition including known boundaries using the Lebesgue measure.

Although the GPE estimates only agree in the limiting value (as the parameter $K\rightarrow N$) with those of the associated histogram estimates of DSE for samples taken from an uniform distribution, it is generally accepted that entropy is not well-defined in the context of a continuous state space, as values are equivalent up to additive constants and unit conversions~\cite{landau2013statistical}. Due to quantization, a well-defined value can be obtained at the quantum level~\cite{partovi1990entropy,penrose2005road,ben2007entropy}, but when dealing with macroscopic scale observations, some level of coarse-graining is required to enable meaningful representation of the state space from a finite sample~\cite{sethna2006statistical}.  Under such an assumption, it was shown that GPE provides a more consistent and informative entropy measure with respect to variation in parameter choice as compared to histogram estimates of DSE, especially when applied to sparse samples from complex distributions~\cite{diggans2022geometric}.  This discrepancy in value will be further explored here in light of the work of Kolmogorov and Sinai in relation to dynamical systems, where we will find support for the claim that the limiting value of histogram estimates are extremely biased and untrustworthy, while those of GPE are more informative and based on theoretical limits (though still biased by sample size).  

Regardless, the ability of GPE to incorporate the impact of meaningful outliers and leverage small samples effectively positioned it as a significant advancement in non-parametric entropy estimation. Furthermore, when the quantile-based partitioning is applied to time series analysis, it was shown to efficiently estimate entropy from small samples and create more ergodic symbolic Markov chain approximations of the underlying dynamics from limited observations~\cite{diggans2022geometric}. Interestingly, the field of symbolic dynamics is also entwined with the production of ergodic state transition graphs and similarly relies on partition entropy in its formulation; this connection will be explored in detail below.

In addition to its being limited to only one dimension, the original quantile-based estimator for GPE struggled to properly account for exactly repeated values.  That failing was shown in~\cite{diggans2023boltzmann} to be related to the failure of histogram estimates of DSE to properly account for outliers, and both of these challenges are now further connected through the lens of Kolmogorov.  While the introduction of the Boltzmann-Shannon Interaction Entropy (BSIE) in~\cite{diggans2023boltzmann} exploited the symmetric relationship of these approaches to define a normalized entropy measure that mitigated these effects to a degree, repeated values can be dealt with more directly within the measure-based framework in order to satisfy the requirement of Kolmogorov's definition that the partition contain no set of measure zero. Through the inclusion of an inverse frequency in the measure used, the discrete probability distribution obtained by the new approach to GPE is more representative of a collection of non-zero local \textit{specific volume} measures. We then further generalize this framework to include alternative choices in measure that can be tailored to various analytical goals.  Thus, geometric partition entropy is now re-imagined in the context of bounded regions in $\mathbb{R}^d$ and presented in a more mature and adaptable way.

Section~\ref{sec:review} provides a review of several foundational concepts for information theory on continuous state spaces, including the definition of Differential Shannon Entropy (DSE) and the associated concept of partition (or metric) entropy, which leads to a discussion of symbolic dynamics and the overall process of symbolization of data from continuous state spaces while incorporating repeated observations. After then providing a definition of Mutual Information (MI) and describing the KSG estimator in detail, Sec.~\ref{sec:GPE} introduces a new definition for GPE, which now both incorporates a pre-processing step that enables the proper incorporation of exactly repeated values and leverages the concept of a Voronoi diagram. Following the correspondence principle, the new definition collapses to a form of the estimator in~\cite{diggans2022geometric} when applied to a set of unique observations with $d=1$.  Several computationally efficient estimators are then defined based on alternative partitioning schemes, which avoid computing the Voronoi diagram.

These approximations are then leveraged to define a new class of estimators for the foundational quantity of Mutual Information (MI) in Sec.~\ref{sec:MI}.  The full power of this new paradigm is observed through the consideration of alternative measures that can highlight the impact of outliers in a variable manner.  Direct comparisons with both histogram estimates of MI and the popular KSG estimator are provided in Sec.~\ref{sec:results} to explore the computational complexity and scaling with dimension, along with an application in synchronization dynamics. We conclude in Sec.~\ref{sec:conclusion} with some remarks on the power of this estimator and suggest some potential adaptations and future applications.

\section{Entropy, Symbolization, and Mutual Information}
\label{sec:review}
The relationship between entropy and mutual information, which is central to modern information theory, is reviewed with a focus on several key concepts with respect to continuous state space estimators that center around the process of discretizing data on a continuum into a partition of a finite set of macrostates to enable a symbolization of the data.  Namely, we introduce the concepts of differential Shannon entropy and the associated partition entropy of Kolmogorov, and then describe symbolization of continuous data including a process for incorporating repeated values into the symbolization using measure-based partitions.  This is followed by an introduction to Mutual Information (MI), including a description of geometric estimators of MI that have recently enabled significant advances in the study of complex systems, data science, and related disciplines.  The following thorough review of related concepts is necessary to properly place geometric partition entropy within the wider context of entropy-based analysis tools and provide motivation for its adoption.

\subsection{Differential Shanno  n Entropy}
Shannon's approach to entropy quantification for a continuous \textit{pdf}, $\rho$, defined over a bounded finite measure state space, $\Omega$, simply took the limit as the number of discrete states goes to infinity, transforming the discrete sum from his theory of communication into the integral form:
\begin{equation}
    H(\rho) = -\int_\Omega{\rho\log_2{(\rho)}}.
\end{equation}
While mathematically reasonable, we will see that fundamental issues can arise when applied to estimation from finite samples.  The integral form has since been termed Differential Shannon Entropy (DSE), and historical approaches to its estimation are primarily divided into two categories: parametric and non-parametric estimators. 

Parametric estimators are based on the assumption of \textit{a priori} knowledge of the underlying distribution, which has the potential to lead to the \textit{petitio principii} fallacy, or ``begging the question". Such estimators can often use the integral form directly, and can also provide reliable entropy estimates for known distributions, \textit{including} a realistic consideration of outliers, but their applicability and effectiveness diminishes when considering samples drawn from distributions of an unknown functional form.  Although many datasets might display certain characteristics that indicate a known functional form, any assumptions of this kind can lead to very biased outcomes. Further, the goal of entropy estimation is to gain insight about an underlying distribution, which, in a sense, can render such estimators paradoxical. Thus, while parametric estimators of DSE are valuable in certain contexts where knowledge about the data is indeed available or assumptions are thoroughly warranted, their utility and accuracy is limited in broader applications. In this work, we will focus on approaches that make no assumptions, setting aside parametric estimators entirely in order to analyze data samples without presupposing any distributional information.  

In the domain of non-parametric estimation of entropy from data, there are two primary approaches that we will consider in detail.  However, there is also a lesser known, but interesting approach to entropy estimation within a totally ordered space, which was relevant to the discussion of the original formulation of GPE, and this method is usually referred to as m-spacings~\cite{pyke1972spacings,hall1984limit,van1992estimating, ghosh2001general}.  Incidentally, the generalization of GPE presented here may also enable a similar approach to generalize that class of estimator to higher dimensions through the use of an alternative choice in measure.  Thus, while we encourage those interested to explore adaptations of m-spacings through the lens of the generalized GPE framework, we will limit our review of common approaches to the two main non-parametric estimators: 1) histogram-based estimators of DSE and 2) $k$-nearest neighbor statistical approaches.  We address the histogram approach first and delay our discussion of the knn methods until after introducing the information measure for which these statistical approaches were developed. 

Histogram estimators are essentially a Riemann sum approximation of the integral form of Shannon's theory, and as their name suggests, they seek to approximate a continuous probability distribution over a bounded region $\mathscr{D}\subset\mathbb{R}^d$ by dividing the region $\mathscr{D}$ into $N$ equal-sized bins (i.e. uniform Lebesgue measure macrostates); the normalized frequencies of observation over these macrostates defines a discrete approximation of the underlying probability distribution ($\rho$), which we denote by the vector ${\bf p}$.  This is used to obtain an $N$-bin estimate of $H(\rho)$ in the discrete symbolic entropy formula
\begin{equation}\label{eqn:DSE}
H({\bf p}) = -\sum_{i=1}^N p_i \log_2{(p_i)}.
\end{equation}

Despite well-known limitations~\cite{silva2012convergence}, this approach to estimating DSE remains the dominant framework across many fields, perhaps due to its intuitive definition; and, given a sufficient volume of data, this approach still makes sense.  However, as the dimensionality of data increases, the required number of observations for a representative sample of the state space grows exponentially.  This phenomenon, known as the ``curse of dimensionality,'' presents a significant challenge to traditional histogram-based approaches in high-dimensional information theory applications, such as those found in the field of machine learning~\cite{mirkes2020fractional}. Thus, for many modern practical applications of information theory, such estimates are infeasible.

Moreover, histogram-based methods come with several other limitations that often lead to a loss of information.  This is particularly true in time series analysis, where the focus becomes the relative frequencies of events rather than their inherent informational value, which may not be uniform~\cite{Cuesta-Frau2019Permutation}. They are also prone to large errors, especially in the case of skewed distributions or those with biased attributes, which are common in real-world data \cite{To2013Entropy-based}.  Some of these failings can be explained through the consideration of the work of Kolmogorov, which we review next.

\subsection{Partition Entropy and Symbolic Dynamics}
Soon after the introduction of Shannon's theory of entropy, the related concept of partition entropy (also known as metric entropy) was first explored by Kolmogorov as a metric invariant in dynamical systems~\cite{kolmogorov1958new}.  Through identification with a Bernoulli process, the adaptation of Shannon's theory to a distribution of measure over a partition of a measure space was described. This metric-based approach to entropy estimation was shown by Y. Sinai to generalize to a Bernoulli scheme with $N$ possible outcomes and probability vector {\bf p} in~\cite{sinai1959concept}, for which the partition entropy is then given by Equ.~(\ref{eqn:DSE}) for this discrete distribution, connecting the histogram estimate of the entropy of a probability density function with the entropy of a measurable partition over a probability space.

The specific case considered by Sinai consisting of a finite partition $\mathscr{A}=\left\{A_1, A_2,...,A_K\right\}$ of a measure space is directly relevant to the formulation of geometric partition entropy, and so we define the partition entropy limited to this case. 

\begin{definition}
\label{def:PE}
    For a bounded set $\mathscr{D}\subset\mathbb{R}^d$, let $(\mathscr{D},\sigma(\mathscr{A}),\mu)$ be the probability space where $\sigma(\mathscr{A})$ is the $\sigma$-algebra generated by a given partition of $\mathscr{D}$ and $\mu$ is taken to be any measure for which $\mu(\mathscr{D})=1$.  If the set $\mathscr{A}=\left\{A_1, A_2,...,A_K\right\}$ satisfies:
\begin{enumerate}[1)]
\item $\displaystyle{\bigcup_{i=1}^K{A_i}=\mathscr{D}}$,
\item $\mu(A_i)>0$ for all $i\in 1,2,...,K$, and
\item $\mu(A_i\cap A_j)= 0$ for all $i\neq j$,
\end{enumerate}
then the \textit{partition entropy} of $\mathscr{A}$ is defined to be 
\begin{equation}
\label{eqn:PE}
H(\mathscr{A}) = -\sum_{i=1}^K \mu\left(A_i\right) \log_2{\left(\mu(A_i)\right)}.
\end{equation}

\end{definition}
With this definition in mind, the basic quantile partition estimate of GPE for a one dimensional data sample, as described in~\cite{diggans2022geometric}, is simply the partition entropy (using the Lebesgue measure) of a particular partition of a bounded interval from which the sample was drawn.  This partition was defined by taking the $i/K$-th quantiles for $i=1,2,...,K-1$ together with the bounds of the state space to form a partition of the interval into $K$ unequal size macrostates. To provide additional context for our generalization to higher dimensions, a more in depth look at the theory behind partition entropy as it pertains to countable partitions will be informative.  

Sinai showed in~\cite{sinai1959concept} that for the case of a general base measure space $(X, \mathcal{B}, \nu)$, where $X$ is a set, $\mathcal{B}$ is a $\sigma$-algebra on $X$, and $\nu$ is a measure, one may consider the relative entropy with respect to some region $\Tilde{X} \subset X$ and a countable partition $\mathscr{B}=\left\{B_1,B_2,...\right\}\subseteq\mathcal{B}$ of $\Tilde{X}$ as long as $\nu\left(\Tilde{X}\right)=1$.  The entropy relative to this subset is then defined by $H(\mathscr{B})$ using the convergent infinite sum of the form of Equ.~(\ref{eqn:PE}), meaning that the treatment of more complex regions $\mathscr{D}$ consisting of unions of multiple disjoint subregions is also valid for GPE estimation as long as a normalized measure can be defined on each component.  In addition, if any sets of measure zero are included in the region $\mathscr{D}$, for instance in a mixed categorical/continuous data setting, GPE can be used on the continuous portion of the data, leaving room for combined estimation strategies using discrete Shannon Entropy on those sets of measure zero, though we have not explored this case at present.

The motivation for the work of Kolmogorov and Sinai was the application of this entropy measure to successive partition refinements under measure preserving automorphisms in order to define a metric invariant of an associated dynamical system, now referred to as the Kolmogorov-Sinai Entropy, $h_{KS}$, of the system.  While the details of this limiting refinement process are not necessary for our purposes, it led to the development of symbolic dynamics, which details a well-defined method of state space partitioning based on the underlying dynamics that converges toward an ergodic state transition graph.  

As noted in~\cite{Strelioff2006How}, ``symbolic dynamics offers a granular view of chaotic attractors that is vital for understanding the inherent unpredictability in these systems,'' and more generally, it has enabled the full application of information theory to the study of continuous dynamical systems.  More specifically, each initial condition for a system can be associated with an infinite sequence of symbols drawn from a set associated with what is called a generating partition of the state space.  The details of symbolic dynamics is beyond the scope of this work, but what is relevant to our argument here is that any two distinct initial conditions of an uncoupled dynamical system will define distinct orbits in the state space that will eventually diverge from each other in their infinite length symbolic representation~\cite{alligood1998chaos,bollt2013applied}. This divergence is indicative of the feature of chaos referred to as sensitivity upon initial conditions, and symbolic dynamics offers unique insight into this property of chaotic dynamics. 

Finally, we point out that while the work of Kolmogorov and Sinai dealt with the action of automorphisms of a dynamical system on a measurable state space, the quantile-based approach to GPE similarly revealed a connection with defining an ergodic representation of the underlying dynamics~\cite{diggans2022geometric}. However, this was achieved by focusing on the relationship between microstates and macrostates as they can be understood in information theory. Namely, each observation can be considered to be an observed microstate, and a set of $K$ macrostates is sought such that the known microstates are as equally represented as possible (essentially an inversion of the physics context of the density of states, where all unobservable microstates are considered to be equally representative of their associated well-defined macrostate).  This approach to macrostate definition leads to a near optimal (limited by the sample variation) partition of the state space for which an ergodic Markov chain model can be constructed in a data-driven way by the observed transitions with no connection to the return map or the underlying dynamics. 

This connection between Kolmogorov-Sinai entropy and geometric partition entropy remained unexplored at the time of the publication of~\cite{diggans2022geometric} and~\cite{diggans2023boltzmann}, however, it provides significant precedent to the concept of GPE more broadly; and when taken together, these two concepts in the symbolization of a continuous state space provide similar insight into the mechanisms of state transitions in complex systems and help bridge the gap between continuous phenomena and their discrete analysis.

\subsection{Symbolization: Discretization of a Continuous State Space}
\label{sec:symbolizing}
Shannon's introduction of entropy in his "Mathematical Theory of Communication Engineering" revolutionized information theory, particularly as it is applied to digital communications through a discrete set of symbols.  The application of Shannon's theory to data obtained from continuous dynamical systems, although common, has always posed significant challenges, primarily centered around the process of discretizing and therefore symbolizing the dynamics.  This is especially true when data are sparse, or the underlying distribution has low probability regions, where it becomes challenging to discretize the continuous state space in a way that reflects the inherent information content of the data. Here, the basic histogram estimator, which relies on the idea of a Riemann sum and thus uses a partition of the domain into equal-measure bins whose importance are weighted by frequencies, leads to several problems.  

One significant issue is the loss of information about the underlying continuous dynamics due to the coarse-graining procedure. Where the work of Kolmogorov and Sinai sought an invariant partition under a given map, the standard histogram approach to entropy estimation naively breaks up the bounded interval into equally spaced subintervals.  Depending on the chosen number of bins, $K$, determining the desired size of the overall symbolic state space, the equal partitioning often oversimplifies the complex variability in such systems.  Thus, the resulting analysis can be sensitive to the choice in $K$, leading to, on the one hand, low fidelity coarse approximations that might not capture critical aspects of the dynamical behavior; while on the other hand, overfitting a particular sample and, the unintentional removal of portions of the true state space from consideration.  

This basic point can be illustrated by considering one of the simplest continuous time chaotic systems.  It can be shown through Finite Time Lyapunov Exponent (FTLE) analysis that the sensitivity to initial conditions in the R\"{o}ssler attractor is concentrated within a small portion of the continuous state space~\cite{diggans2022hierarchy}. In order to represent enough detail in this smaller region while using equal-sized partitions, the overall number of states required increases rapidly, resulting in an explosion in the sample size required for any useful frequency-based analysis.  This type of simplification through equi-partitioning can significantly affect entropy estimates, and assumes that the only measure of informativity for a macrostate is the frequency of observation.  This assumption may be valid under random sampling from a distribution, but when this approach is applied to dynamical systems, this is decidedly not the case, resulting in skewed representations of the system's actual behavior and potentially leading to inaccurate conclusions about information content.

Furthermore, the sample size required to effectively represent low-probability regions in frequency-based methods can be prohibitively large, which often leads to the use of histogram estimates in which some bins have a frequency of zero.  One of the conditions in the definition of partition entropy is that the partition $\mathscr{A} = \left\{ A_1,A_2,...,A_K \right\}$ must be able to generate a $\sigma$-algebra on the set $\mathscr{D}$ (i.e. be a nonempty collection of subsets of $\mathscr{D}$ that are closed under compliment, countable unions, and countable intersections). The frequency-based histogram obtained from a data set can be viewed through the lens of measure theory within the probability space $(\mathscr{D},\sigma(\mathscr{A}),\#_X)$, where the set counting measure $\#_X(A)$ is defined as giving the number of elements from the discrete set $X$ found within the argument $A\subseteq \mathscr{D}$. Under the usual equi-partitioning of a histogram estimate, if a bin $A_i$ has frequency zero, that implies $\#_X(A_i)=0$. Since $A_i\neq\emptyset$ but has measure zero, it should be removed from the sigma algebra and a subset $\Tilde{\mathscr{D}}$ should be considered after removing this bin.  This describes exactly what occurs in the computation of Equ.~(\ref{eqn:DSE}), where zero frequency bins are ignored in the sum.  

This point is central to understanding the failures of histogram estimates of DSE as the set of utilized symbolic states may not be a $\sigma$-algebra on $\mathscr{D}$ if $\cup_i A_i\neq\mathscr{D}$, even if $\cup_i \#_X(A_i)=\#_X(\Tilde{\mathscr{D}})=\#_X(\mathscr{D})=1$. The effective deletion of the empty bins (which is the common practice) is basically ignoring part of the dynamics-occupied geometry, which assumes the system will never experience a state transition to this geometric space.  This is a major pitfall of histogram estimators of DSE, but it also explains why DSE works well when the data are densely distributed over the region.  However, even in such ideal cases, there is another assumption that doesn't practically hold either, which is that a near-uniform density exists within each state of the partition. The best theoretical scenario for such an assumption to hold is if ($1<< k\leq N\rightarrow\infty$). 

Although not as problematic, this second assumption is true for GPE as well, and can be used as an argument for larger values of the parameter $K$ in its estimation. Importantly though, where larger parameters ($K$) in DSE converge to the value $\log_2{(N)}$, this is not true for GPE.  In contrast to histogram DSE estimators, the use of quantiles in the original formulation of GPE prevented the unintentional removal of any of the available state space; and furthermore could incorporate additional unobserved regions of a known larger state space.  This feature along with the theory supporting KS entropy, lends credence to the claim that the limiting value of the GPE estimator that is observed as the number of macrostates, $K$, approached the sample size $N$ is more informative (though still biased by $N$) than the limiting values of histogram estimates, which are strongly biased toward $\log_2{(N)}$.

Similar to the problem that an empty bin poses in histogram estimations of DSE, there is a related challenge in the original discretization process for GPE, which is the proper inclusion of (nearly) repeated values.  Even if a phenomena is assumed to be occurring on a truly continuous state space, many applications will inevitably involve some limit on the fidelity of measurement instruments and/or error tolerances, including machine precision of computation before ever reaching any concerns at the Planck scale.  This will necessarily require the consideration of repeated values within a dataset.  While frequency-based histogram estimates struggle to appropriately include the impact of outliers due to empty bins, they have no problem including the impact of repeated values.  On the other hand, any geometric approach, including both knn-based approaches and GPE, must address this challenge. The impact on knn-based methods can be mitigated to a degree for small numbers of repeated values by using a larger $k$ value, but this strategy fails when dealing with dynamics that ever display even nearly periodic behavior.  

And, while GPE can deal effectively with near periodic dynamics, the quantile-based estimator of GPE from~\cite{diggans2022geometric} fails in the presence of even a small number of exactly repeated values.  This concern was, in part, the motivation for the introduction of the Boltzmann-Shannon Interaction Entropy, and so a more complete discussion on this point was presented in~\cite{diggans2023boltzmann}. This failing of the original quantile-defined GPE estimator can also be understood under the requirements of KS entropy; i.e. in the presence of repeated values, quantiles may coincide resulting in zero Lebesgue measure bins.  The new approach to estimating GPE in $\mathbb{R}^d$ will now address this important challenge head on by requiring a tolerance for the inclusion of repeated values (which can default to just above machine precision). This repetition of the observed state must then be accounted for in the computation of GPE through an appropriate weighting, which can be effectively defined as a measure-based abstraction of the concept of specific volume as will be described in section~\ref{sec:GPE}. 

\begin{figure}
    \centering
    \includegraphics[width=7cm]{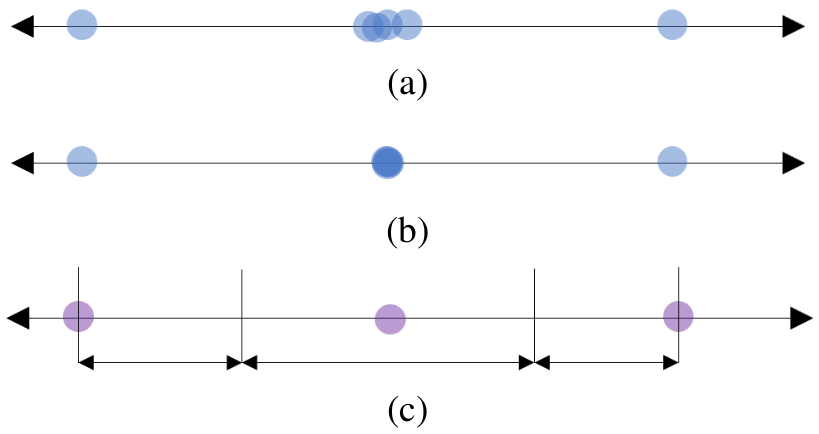}\\
    \caption{(a) and (b) show two different sets of measurements of the same phenomena as recorded by devices that have larger and smaller error tolerance respectively. It is not clear based on the known parameters of the measurement devices that the central clusters should be separated into a collection of small macrostates, and in the absence of such information, we must assume that these measurements belong to the same coarse-grained macrostate. (c) illustrates the pre-processing alteration to the dataset that would be allowed in GPE, where the continuous values are binned within known tolerances, but left defined on the continuum.  Ideally, in GPE for a particular course-graining, each macrostate will be equally represented by the data, but this is not always possible, e.g., computers generally operate at machine precision of $\epsilon\approx 10^{-15}$, meaning some level of discretization is always necessary before course-graining.}
    \label{fig:uniqtol}
\end{figure}

To illustrate the main idea underlying the pre-processing step that will enable proper incorporation of repeated values, two data sets are shown in Fig.~\ref{fig:uniqtol} (a) and (b) that may result from two different measurement devices having differing fidelity, but recording the same underlying phenomena.  Choosing the correct error tolerances for the respective measurement devices should result in both sets being partitioned, as shown in (c), where the center macrostate now represents four nearly repeated measurements.  This type of pre-processing can avoid the definition of trivial (zero width) or otherwise spurious (within tolerance) macrostates based on the quantiles between noisy measurements of what is better classified as the same underlying state.  Following this type of pre-processing of a data set, we will begin the analysis with a potentially reduced number of unique observed microstates, still defined on the continuum, along with an associated frequency of microstate observation vector. 

\subsection{Marginals Spaces and Mutual Information}
\label{subsec:MI}
The interplay of entropy and Mutual Information (MI) is a foundational relationship of modern information theory. The entropy of a data set serves as a measure of uncertainty, quantifying the spread of a sample and therefore providing a measure of predictability in the phenomena from which it was recorded.  Mutual Information builds upon this basic concept by quantifying the extent to which knowledge of one variable reduces the uncertainty (or equivalently increases the predictability) about another. Of particular interest is its application to the study of synchronization, where an increase in MI between two time series is observed when the initial conditions and chosen coupling strength leads to synchronization.

\begin{figure}
    \centering
    \includegraphics[width=7cm]{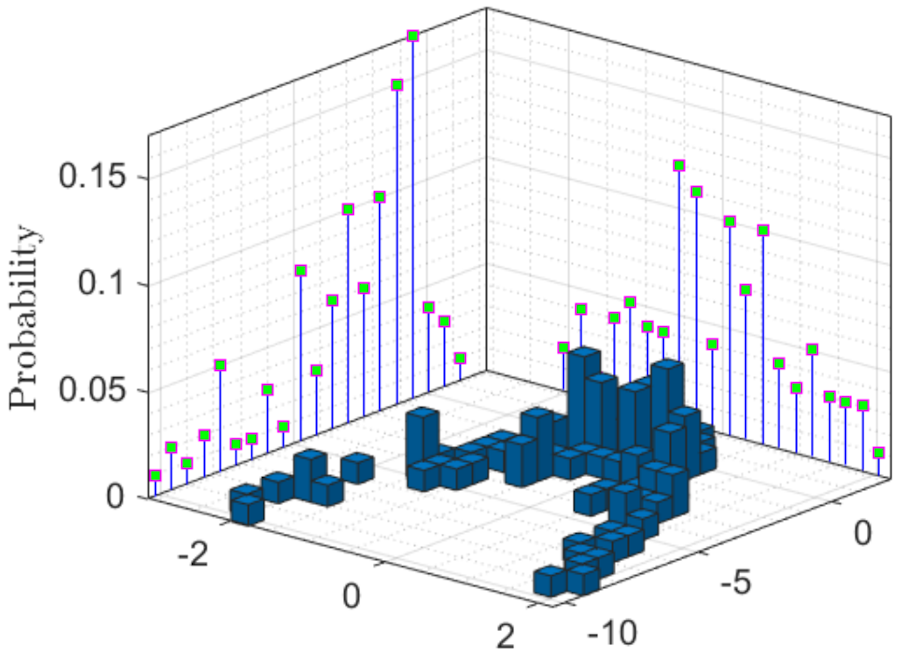}
    \caption{A histogram estimate for a probability distribution from a simple data set $(X,Y)\in\mathbb{R}^2$; the joint entropy, $H(X,Y)$, would be computed from this histogram in the joint space using Equ.~\ref{eqn:DSE}, whereas the marginal entropies $H(X)$ and $H(Y)$ would be computed using Equ.~\ref{eqn:DSE} on the one dimensional histograms shown on the vertical planes.}
    \label{fig:marginalGeo}
\end{figure}

In order to define MI, we must first introduce several related concepts that are associated with high dimensional data sets obtained through the concatenation of several lower dimensional state spaces. For example, if $X\subset\mathbb{R}^{d_1}$ and $Y\subset\mathbb{R}^{d_2}$ are equal length sets of observations in $d_1$ and $d_2$ dimensions respectively, then we can consider the data set $(X,Y)$, consisting of observations in the joint space, $\mathbb{R}^{d_1+d_2}$, where the pairwise associations of observations in $X$ and $Y$ are considered an important feature of the joint data set.  One can then estimate the \textit{joint entropy} of this data in the concatenated space, denoted as $H(X,Y)$, simply by using a $d=d_1+d_2$-dimensional histogram with Equ.~\ref{eqn:DSE}; although here is where the sample size required for accurate estimation often prohibits the use of such estimators unless $d$ remains low.  To illustrate, Fig.~\ref{fig:marginalGeo} shows a histogram representation of a probability distribution for a low dimensional data set in the case of $d_1=d_2=1$, where the histogram approximations for the marginal distributions, obtained by projecting the data set $(X,Y)$ onto the marginal spaces $X$ and $Y$, are shown on the vertical panels.  The marginal entropies, $H(X)$ and $H(Y)$, are estimated from these projected histogram approximations of the marginal probability distributions of the data set using Equ~(\ref{eqn:DSE}). 

Another important and related concept is the \textit{conditional entropy}.  The entropy of $Y$ conditioned on $X$ is denoted by $H(Y\mid X)$ and seeks to quantify the remaining uncertainty in the random variable $Y$ (the amount of information required to encode $Y$), given that we know the value of another random variable $X$. In other words, $H(Y \mid X)$ tells us how uncertain $Y$ is if $X$ is known. If $X$ and $Y$ are independent, knowing $X$ does not provide any information about $Y$, and $H(Y \mid X) = H(Y)$. Conversely, if $X$ completely determines/describes $Y$ (though they are not necessarily equal), then $H(Y \mid X)=0$ because there is no uncertainty left in $Y$ once $X$ is known.

Using these concepts, the Mutual Information (MI) between two random variables $X$ and $Y$, can be defined in several ways including:
\begin{eqnarray}
    I(X;Y) & = & H(X) - H(X \mid Y) \notag \\
           & = & H(Y) - H(Y \mid X) \notag \\
           & = & H(X)+H(Y)-H(X, Y). 
\label{equ:MI}
\end{eqnarray}

Based on the theoretical bounds of the component entropies in these definitions, it should be clear that the mutual information is itself bounded, i.e. $0\leq I(X;Y) \leq H(X)$, and in particular, it should never be negative.  

However, the joint and marginal entropy estimates for data in continuous state spaces are often biased in different ways, e.g. toward the number of bins with non-zero measure, which may differ due to the sample $(X,Y)$ being more sparse in the joint space with respect to its dimensionality as compared with the marginal distributions.  This particular indication of the curse of dimensionality has been known to result in negative estimates of mutual information as the joint entropy is artificially inflated by sparsity and the larger number of zero frequency bins. 

\subsection{Knn Estimators of Mutual Information}
A great leap forward in entropy estimation came in the form of the $k$-nearest neighbor (knn) mutual information estimator introduced by Kraskov, St\"{o}gbauer, and Grassberger in~\cite{ksg2004estimating}, which is now often referred to as the KSG estimator.  The KSG mutual information estimator relies on the Kozachenko-Leonenko (KL) estimate for Shannon entropies that was introduced decades earlier in~\cite{kozachenko1987sample}.  Specifically, it calculates the distance to the $k^{th}$ nearest neighbor in the joint space and by assuming that each point has equal representation in the joint space, it uses this distance to determine the number of neighbors in the marginal spaces. By averaging these local estimates, the KSG estimator effectively captures the mutual information between the variables according to the last form of Equ.~(\ref{equ:MI}).  

Although this approach has been celebrated in recent decades for its application in data-driven analysis of complex systems, we will show that its success is largely due to statistical averaging, in which all outlier impacts are negated.  Since the introduction of the KSG estimator, there have been various improvements to the basic approach through the incorporation of more sophisticated geometric approximations, including~\cite{lord2018geometric}. However, the KSG estimator remains the most popular MI estimator for data in high dimensions.  This is likely due to its ease of application to estimates of more complex measures including Conditional Mutual Information (CMI). Although we will not be concerned with CMI here, we restrict our focus to the KSG approximation as the most popular representative of this class of geometric estimators.

\begin{figure}[ht!]
    \centering
    \includegraphics[width=8cm]{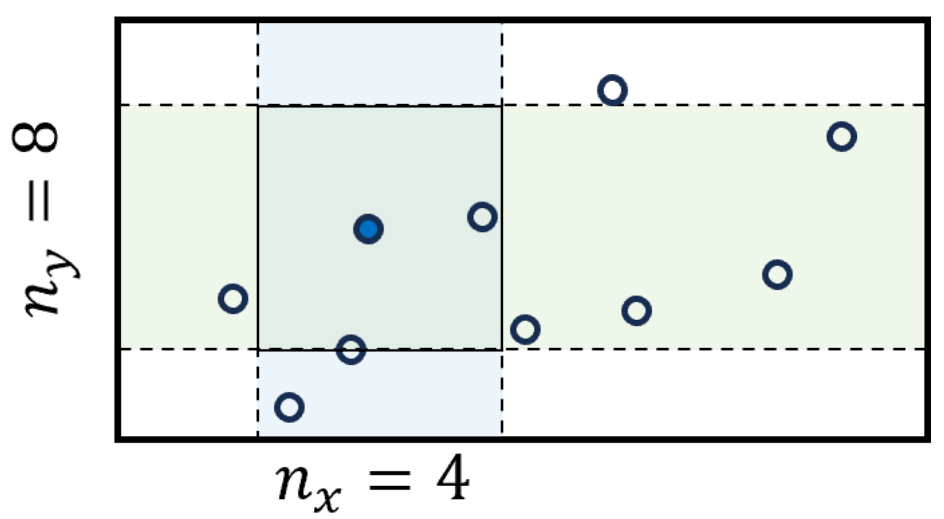}
    \caption{A representative illustration of how the marginal probability estimates are computed for each data point in the KSG estimator; for example when $k=2$, a square neighborhood (an open ball under the Chebyshev distance) is defined by the $k$-th nearest neighbor, and the number of data points that fall within this neighborhood in each of the marginal spaces are counted to define $n_x$ and $n_y$.}
    \label{fig:knn}
\end{figure}

The KSG estimator commences by identifying the Chebyshev distance, which we will denote as $\nu$, to the $k$-th nearest neighbor from each data point.  For each point, a box of width $2\nu$ centered around the point is defined, e.g., in a two-dimensional (2D) context, this is a square of area $(2\nu)^2$ centered around the data point, as represented in Fig.~\ref{fig:knn} for parameter $k=2$.  The frequency of observations in each of these square neighborhoods in the joint space is then $k$ by definition, but the estimator calculates the marginal probabilities (frequency) based on the projections of that square onto the marginal spaces, e.g., points within the dashed rectangles projected from the square in Fig.~\ref{fig:knn}.  Defining the marginal probabilities associated with each point as $n_x$ and $n_y$ (frequencies within the marginal spaces $X$ and $Y$, respectively, that lie within the constructed joint space around the point), the mutual information between two random variables $X$ and $Y$ is then estimated as:

\begin{equation}\label{eq:ksg}
    I(X;Y)=\psi(k)+\psi(N)-\left\langle\psi\left(n_x+1\right)+\psi\left(n_y+1\right)\right\rangle,
\end{equation}
where $\psi$ is the digamma function ($\psi(z)=\frac{\mathrm{d}}{\mathrm{d} z} \ln \Gamma(z)=\frac{\Gamma^{\prime}(z)}{\Gamma(z)}$), $k$ is number of nearest neighbors (a parameter), $N$ is the sample size, and $\left\langle \ldots \right\rangle$ indicates an average over all $i \in[1, \ldots, N]$.

What often goes unstated in the literature is that the argument of this final term is itself an approximation, meaning the term $\left\langle\psi\left(n_x+1\right)+\psi\left(n_y+1\right)\right\rangle$ only represents the average contributions from the marginal entropies.  From a frequency viewpoint, this statistical approximation translates to the assumption that the data points are uniformly probable within the joint space, allowing for a cancellation of the geometric bias terms that appear in the marginal entropy estimations in~\cite{kozachenko1987sample}, which resulted from the non-uniformity of the density over the sample.  Since this geometric bias depends on the $k^{th}$ neighbor distances in the marginal spaces, KSG considered the approximation of averaging over all realizations of $k$ to be able to cancel the bias leading to the simplified formula in Equ.~(\ref{eq:ksg}). 

Considering the law of large numbers, this approximation becomes exact only in the limit of large samples, i.e. $N\rightarrow\infty$.  Thus, these underlying assumptions can lead to the requirement of large sample sizes to detect the MI between highly dependent samples, but are also responsible for the incredibly stable, yet potentially problematic estimates provided by the KSG estimator.  Due to the overall stability of KSG estimates with respect to changes in the parameter $k$, in all applications here, we will use a commonly accepted good parameter choice of $k=5$ nearest neighbors.  

Due to these often unstated assumptions, it is informative to note that a failure is seen in the simple act of computing the mutual information between a data set $X$ and itself.  Although this would seem to provide a KSG estimate of the entropy of $X$, since $H(X) = I(X;X)= H(X)-H(X\mid X)$ since $H(X\mid X)=0$, this estimate ends up being meaningless, as shown by comparing the values obtained for widely different data sets in Sec.~\ref{sec:GPE}. 

This statistical smoothing has been celebrated as something that ostensibly enhances the robustness of KSG. However, as described in the introduction, while this averaging method seems to offer resilience, it can be misleading in practice. It doesn't truly increase the estimator's resistance to outliers; rather, it simply mitigates their impact by spreading it across the entire dataset. This treatment of outliers is particularly concerning in datasets associated with dynamics, where these atypical points may hold significant, non-trivial information, as the averaging process could obscure crucial insights. For example, when estimating MI over a synchronization process, the role of the initial conditions are completely lost as the transient period gets treated as outlier data in the joint space.

Moreover, this approach to achieving 'robustness' comes with a significant trade-off in terms of computational efficiency. The process of sorting through the nearest neighbors of a large number of data points necessitates increased computational resources and time, particularly as the dimension of the sample grows. Thus, in scenarios involving substantial datasets, increasingly common in fields like data science and machine learning, the KSG estimator's computational burden may render it impractical, which limits its applicability in these areas.

In summary, this class of geometric estimators, which includes the KSG method, still rely heavily on large datasets for consistent estimation, and as such, often require substantial computational resources. Furthermore, like histogram estimators, they face challenges in differentiating between sparsely sampled regions and completely disallowed regions of the state space (in this case due to averaging). Despite these limitations, geometric estimators have enabled significant progress in the study of more stable dynamical systems by providing a consistent MI estimator.  As such, their utility in complex, high-dimensional spaces is still subject to ongoing research and development.

\section{Generalizing Geometric Partition Entropy}
\label{sec:GPE}
A new presentation of geometric partition entropy is now provided that can be applied to data samples drawn from distributions whose domain is a bounded region of $\mathbb{R}^d$.  Although different in its theoretical foundation from the previous approach, the new definition collapses to the one-dimensional estimator described in~\cite{diggans2022geometric} when it is applied to non-repeating samples with $d=1$ and $K\mid N$.  In addition, a frequency-based weighting is now incorporated with the geometric measures of the $K$ macrostates making the basic representation more akin to measures of local specific volumes.  This not only addresses the impact of uneven representation caused by the pigeonhole principle when $K\nmid N$, but it also allows for a more elegant handling of exactly repeated values.

We begin in Subsec.~\ref{subsec:prelims} by emphasizing the necessity of a well-defined boundary of finite measure for the domain $\mathscr{D}\subset \mathbb{R}^d$ to ensure meaningful probabilities; then, go on to describe the pre-processing step in more detail that ensures the appropriate handling of repeated values.  Once these preliminary details are discussed, the Voronoi-based definition of GPE is provided in Subsec.~\ref{subsec:GPE}.  Unfortunately, any actual estimates based on this theoretical definition are computationally expensive, especially in higher dimensions; and furthermore, the use of irregular partitions in the joint space leads to inherent challenges for identifying marginal probability distributions. Thus, we follow the Voronoi-based definition in Subsec.~\ref{subsec:Approx} with several approximation techniques that address these challenges directly through leveraging alternative partitioning strategies and adapted measures effectively.  

After a comparison of these estimators with traditional approaches in Subsec.~\ref{subsec:compare}, several of the approximation methods enable the subsequent definition of a class of powerful and efficient mutual information estimators in Sec.~\ref{sec:MI}, which are able to incorporate informative outliers in a dynamic and versatile way.

\subsection{Preliminaries}
\label{subsec:prelims}
\subsubsection{Ensuring a Bounded Region}
A bounded (finite measure) region is required for the definition of meaningful probabilities with respect to volumes.  In scenarios that lack explicit boundary definitions, approximations of effective limits can significantly influence the estimate of GPE.  This is not seen as a failure of GPE, but rather an important feature, specifically, because one goal of this approach is to properly include the impact of outliers, especially when they might be informative to understanding switching and/or chaotic dynamical systems.  

Ordinarily, the convex hull of a dataset serves as a reliable boundary in many applications, but not necessarily so here.  The convex hull uses actual data points to define the boundary, meaning the regions around those particular observations will be underrepresented unless the observations are in fact extremal values within the state space.  At the same time, the potentially large regions between these boundary points might become overly emphasized if they are not reasonably expected to be in the true state space.  The use of an alpha shape can omit the large unobserved sections of the state space and can be generally advantageous, but these still use data points to define the bounds, likely limiting the proper inclusion of outlier impacts.  

As such, incorporating topological data analysis or relying on other data-driven techniques, such as singular value decomposition or the use of ghost points within a Delaunay Triangulation, can offer a more sophisticated, data-informed approach.  However, in an effort to avoid distraction from the main concepts being introduced, we will proceed under the assumption that a known boundary of $\mathscr{D}$ has been provided.  That being said, some of the approximation techniques presented are (or at least can be) utilized in the absence of a well-defined boundary, but the impact of this decision on the resulting estimates should be explored and must be understood. 

\subsubsection{Handling Repeated Observations within Tolerance}
In addition to a well-defined boundary of a measurable state space, we must also address the known problem from the previous approach of repeated values.  In theory, infinitely accurate observations made on a continuum will result in a zero percent chance of obtaining exactly repeated values.  However, in practice, this is never truly the case.  Putting aside more foundational reasons, e.g., machine precision or even the Heisenberg uncertainty principle, there is more often some known measurement error that effectively limits the accuracy of observations even further.  For instance, there is often little interest in considering differences in the outside temperature beyond a single degree, and there is almost no interest in the minor fluctuations around some average due to transient air flows.  

Thus, any analysis of this type must begin by first defining a tolerance $\epsilon$ that determines when observations in the state space should be considered effectively the same, i.e. as a repeated observation of the same microstate.  This initial pre-processing step, importantly, does not discretize the state space, but rather redefines the data itself within the acceptable limits of measurement accuracy in order to prevent the eventual definition of macrostates with sizes below the limits of measurement tolerance. 

Thus, for a chosen tolerance $\epsilon$, two points $x_i, x_j$ in a dataset $X\subset \mathscr{D}$ consisting of $N$ points can be considered the same observable microstate (within tolerance) whenever $$\Vert x_i-x_j \Vert_\infty \leq \epsilon.$$  This means that we might end up with $M<N$ unique observed microstates along with an associated vector of the repetition number for each state. Importantly, this prevents any element of a partition of the space from having measure zero, which is required by the formulation of partition entropy put forward by Kolmogorov and Sinai and discussed in Sec.~\ref{sec:review}.

\subsection{The Voronoi-Based Geometric Partition Entropy}
\label{subsec:GPE}
Thus, given a set of observations $X=\left\{x_n\right\}_{n=1}^N$ within a bounded (finite measure) region $\mathscr{D}\subset\mathbb{R}^d$, a parameter $K$, and a tolerance $\epsilon$, we first identify any observations $x_i\in X$ such that there exists at least one $x_j\in X$ with $\Vert x_i-x_j \Vert_\infty \leq \epsilon$. A greedy merging process can use centroids of subsets to obtain a set of unique microstates (within a tolerance of $\epsilon$) along with a corresponding repetition number.  After this initial pre-processing step is complete, there are $M\leq N$ unique observed microstates defined on the continuous space $\mathscr{D}$ with an additional frequency of repetition vector of length $M$. 

Similar to the approach in~\cite{diggans2022geometric}, we now seek a partition $\mathscr{A}=\left\{A_1,A_2,\ldots,A_k\right\}$ of the state space $\mathscr{D}$ such that each subset $A_i$ contains roughly the same number of \textit{unique} observed microstates while ensuring that $\cup_i {A_i}=\mathscr{D}$.  This was accomplished in one dimension through the computation of the $i/K$-th quantiles for $i\in 1,2,\ldots,K-1$, taken together with the bounds of the state space (but alternatively could be done through a midpoint partitioning procedure as well).  This step in the process, however, in and of itself, is now a challenging problem with no clear optimal solution due to a lack of a total ordering of the data in higher dimensions. 

A path forward is identified by first considering the special case of $K=M$ (the number of unique microstates), which does have a unique optimal solution provided by the Voronoi diagram of the set of unique observations in $\mathbb{R}^d$ intersected with the boundary of the region $\mathscr{D}$. This results in a geometric partition $\mathscr{V}=\left\{v_1,v_2,...,v_M\right\}$ of the state space $\mathscr{D}$, where each Voronoi cell $v_m\in \mathscr{V}$ represents the region of the state space that is closest to the unique observation, $x_m$.  One might say that each Voronoi cell, $v_m$, represents a macrostate of the continuous state space for which the single observed microstate, $x_m$, is considered the best representative.  This particular partition, however, corresponds to the assumption that the data set is, in fact, an optimally chosen independent and identically distributed (i.i.d.) sample from the true underlying distribution.  This rather unrealistic assumption can be indicative that using $K=M$ overfits the estimate of the underlying distribution to the particular sample, and so, choosing $K<M$ generally provides a more robust estimate.

In the jargon of measure theory, we consider the probability space $(\mathscr{D},\sigma(\mathscr{V}),\mu)$ with the $\sigma$-algebra generated by the partitioning set of truncated Voronoi cells ($\mathscr{V}$), which satisfies the assumptions in Def.~\ref{def:PE}, and the measure $\mu$ is taken to be 
\begin{equation}
\label{equ:measure}
    \mu(A) = \frac{N}{m(\mathscr{D})} \frac{m(A)}{\#_X(A)},
\end{equation}
where $m(\cdot)$ is the Lebesgue measure and $\#_X(\cdot)$ is the same \textit{set counting measure} described in Subsec.~\ref{sec:symbolizing}, meaning $\#_X(A)$ gives the number of elements of the data set $X$ that fall within the geometric region $A$.  The combined measure $\mu$ is representative of the concept of a local measure of \textit{specific volume}, i.e. an inverse density.  Due to the definition of the Voronoi diagram, we need not worry about any sets for which $\#_X(A)=0$, and the multiplicative constant, $\frac{N}{m(\mathscr{D})}$, is chosen to enforce the probability constraint so that $\mu(\mathscr{D})=\sum_{m=1}^M{\mu(v_m)}=1$.  

Due to obvious challenges that can arise with the Lebesgue measure in higher dimensions, e.g. co-linear points etc., some of the approximation techniques will employ an alternative for $m(\cdot)$, but an alternative must satisfy the definition of a measure. Regardless of the chosen geometric measure, it is important to note that $\#_X(\cdot)$ will properly include any repeated values. 

In order to not overfit the estimate to the particular data sample, we will want to consider partitions of $\mathscr{D}$ with $K<M$ partition elements.  Due to the lack of a total ordering in higher dimensional spaces, it is no longer clear which observations should be grouped with what set of nearby neighbors so that $\mathscr{D}$ is partitioned into $K$ subregions having nearly equal representation among the observed microstates.  Importantly, despite this level of uncertainty, it is clear that all such partitions can be represented by members of the $\sigma$-algebra generated by $\mathscr{V}$, i.e. $\sigma(\mathscr{V})$, through unions of the various $v_m$ within each cluster.  Generally, clustered groups should be formed by geometrically close neighbors in a way that does not result in any of the observations being left out, but this is indeed a problem without a single well-defined solution.  However, the benefit of larger values of $K$ means that even greedy clustering of several nearby observations can provide an effective clustering. 

In general, a Constrained K-Means (CKM) clustering approach can be used to great effect, and we include several examples; however, algorithmic convergence is not guaranteed for smaller $K$ values and this approach can struggle under the specific narrow constraints that each cluster include a minimum of $\lfloor M/K \rfloor$ observations. Thus, we wish to encourage those interested to devise algorithmic solutions to this constrained clustering problem for accurate GPE estimation in particular.  For this purpose, we have developed a simple deterministic quantile-based Recursive Prime-Factor Splitting (RPFS) algorithm that provides a reliable and repeatable equitable clustering. The details of this algorithm will be provided in the next subsection where it will enable several approximation techniques. 

Regardless of how a set of equitable clusters are obtained, the estimation of GPE proceeds by fusing the corresponding Voronoi cells of the cluster members from each group into larger subregions (i.e. supercells), which importantly are members of $\sigma(\mathscr{V})$ with positive measure under $\mu$. The GPE estimate is then computed for the set of supercells using Equ.~(\ref{eqn:PE}) with the measure~(\ref{equ:measure}).

\begin{figure}[ht!]
    \centering
    \includegraphics[width=8cm]{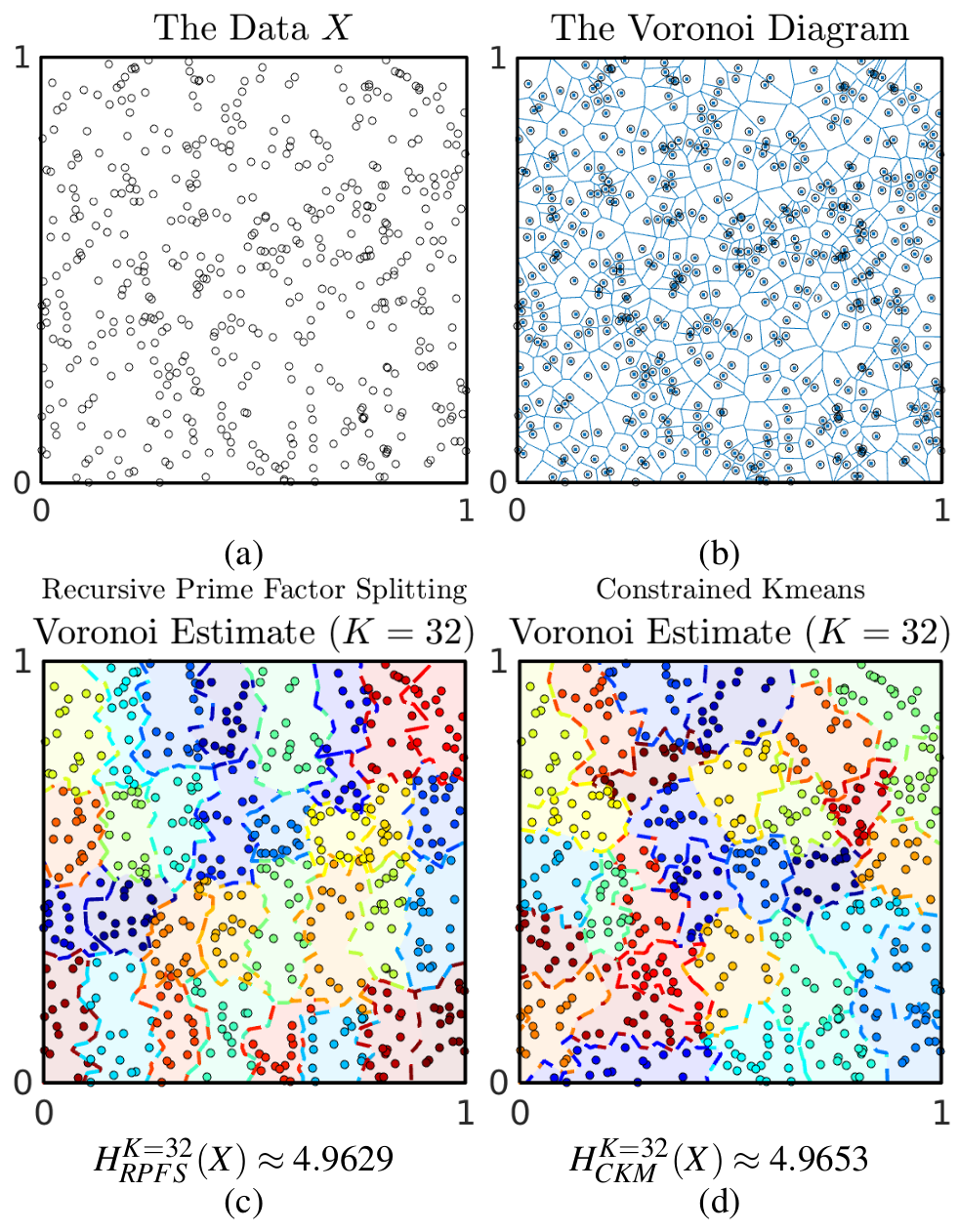}
    \caption{(a) Given a sample of $N=500$ unique values (within a tolerance of $\epsilon=10^{-5}$) lying within a bounded region $\mathscr{D}$, 
 (b) the Voronoi diagram of the data set intersected with the boundary of the region $\mathscr{D}$ provides the optimal partition of $\mathscr{D}\subset\mathbb{R}^d$ for the choice of $K=N$ regions; coarse-grained estimates of GPE can be obtained for $K=32$ macrostates by fusing nearly equally represented groups obtained by (c) RPFS or (d) CKM into a partition of supercells (indicated by shared color) before computing the local specific density measures and using Equ.~\ref{eqn:PE} to estimate the GPE.}
    \label{fig:UniformVor}
    \end{figure}

Figure~\ref{fig:UniformVor} (a) shows a sample of $N=500$ points drawn from a uniform distribution along with (b) the Voronoi diagram of this data set (intersected with a bounding box of $[0,1]\times [0,1]$); whereas panels (c) and (d) show the coarse-grained clustering of this set (by color) resulting from the RPFS and CKM algorithms respectively, both using the parameter value of $K=32$.  The corresponding GPE estimates are $H_{RPFS}^{K=32}\approx 4.9356$ and $H_{CKM}^{K=32}\approx 4.9640$.  

Since this is a uniform distribution, these estimates are expected to coincide in value with the associated histogram estimate of DSE using Equ.~\ref{eqn:DSE}, which gives $H_{DSE}^{K=32}\approx 4.9464$. Finally we note that the supercells should ideally be path connected, but they need not be in order to provide a useful estimate.

\begin{figure}
    \centering
    \includegraphics[width=8cm]{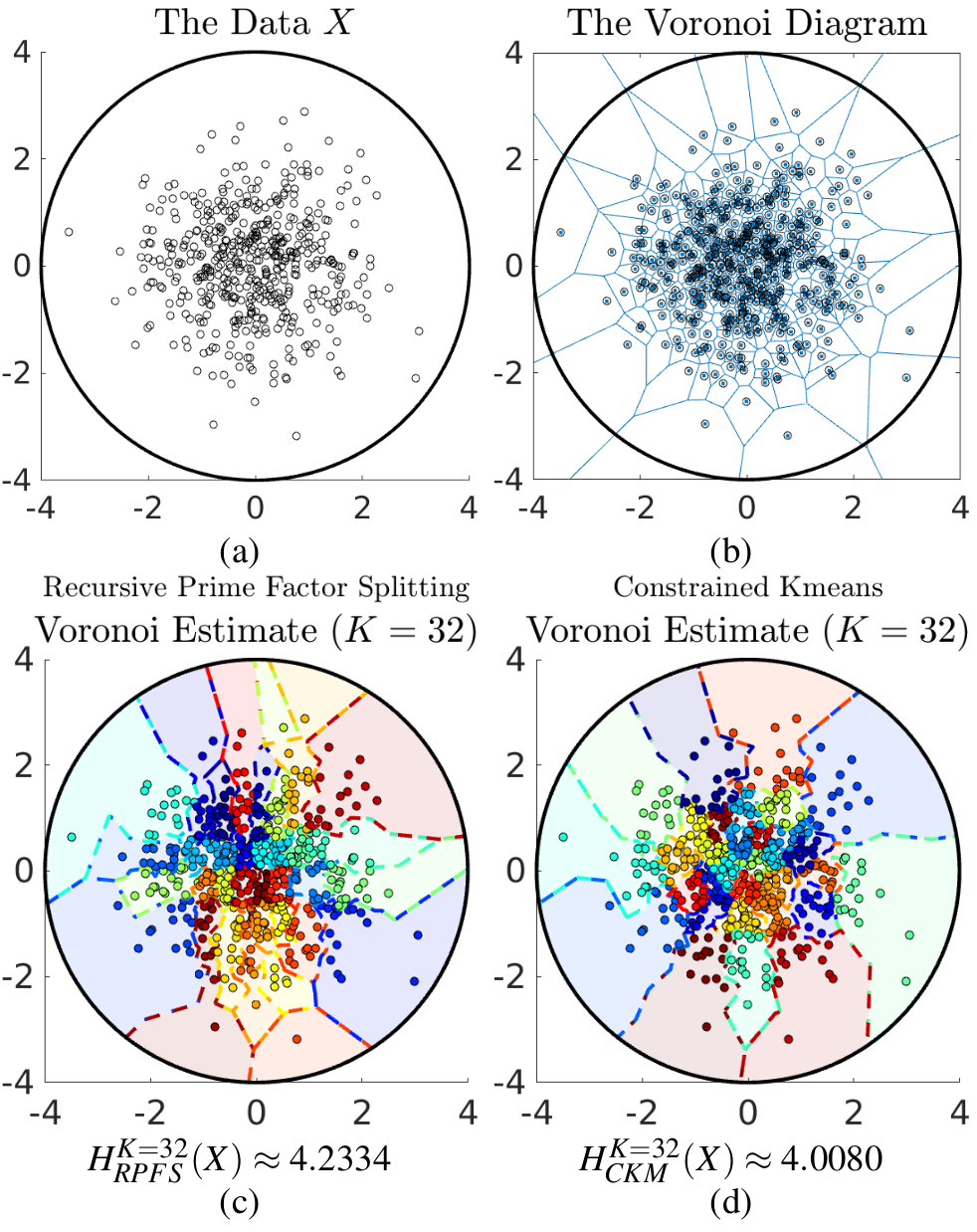}
    \caption{(a) Given a sample of $N=500$ unique values (within a tolerance of $\epsilon=10^{-5}$) lying within a radially bounded region $\mathscr{D}$, 
 (b) the Voronoi diagram of the data set intersected with the boundary of the region $\mathscr{D}$ provides the optimal partition of $\mathscr{D}\subset\mathbb{R}^d$ for the choice of $K=N$ regions.  A coarse-grained estimate of GPE can be obtained for $K=32$ by fusing (near) equally represented groups into supercells (indicated by shared color) before computing the local specific density measure of these supercell macrostates; (c) and (d) show two examples of supercell partitions obtained with the RPFS and CMK algorithms respectively.}
    \label{fig:NormalVor}
\end{figure}

Similarly, we consider a random sample of $N=500$ points drawn from a standard multivariate normal distribution with an imposed radial boundary representing a $4\sigma$ radius. Figure~\ref{fig:NormalVor} (a) shows this data sample together with (b) the basic Voronoi partition obtained by intersection with $\mathscr{D}$, and (c) and (d) each showing the RPFS and CKM clusterings of the data into $K=32$ supercells, respectively (as indicated by shared color).  The GPE estimates obtained from this sample by these two clusterings are $H^{K=32}_{RPFS}\approx 4.2334$ and $H^{K=32}_{CKM}\approx 4.0217$ respectively, which now deviate from the DSE estimate of $H_{DSE}^{K=32}\approx 3.3386$ since many of the $K=32$ equal measure bins, which get defined over a rectangular bounding box of $[-4,4]^2$ that contains $\mathscr{D}$, will have zero representation.  Further, even if the parameter $K$ is adjusted such that there are $32$ macrostates that lie within the bounded region $\mathscr{D}$, there is no way to incorporate the impact of unsampled portions of the known state space. This results in not only a bias toward $\log_2{(K)}$ for a lower effective $K$ value, but also innaccuracies due to the handling of outliers.

Although Voronoi diagrams provide a theoretically accurate definition of GPE, their use presents significant computational challenges.  The construction and storage of Voronoi diagrams necessitate exponential costs, especially when adapting to uncertain spatial data where the development of Uncertain-Voronoi diagrams can be of use~\cite{Xie2012UV-diagram}. In addition to the cost of constructing the diagram itself, the direct computation of most geometric measures of the resulting irregular shapes can also become problematic in high dimensions using the Lebesgue measure. 

To this point, other choices for the probability measure can provide some advantage when used in coarse-grain estimation, but keep in mind the requirements of Kolmogorov, namely that $\mu$ must be a measure.  For instance, the use of the Chebyshev norm in place of $m(\cdot)$ enables faster computation in high dimensions, but it does not satisfy the definition of a measure in $\mathbb{R}^d$ and leads to inaccuracies.

Furthermore, for the eventual goal of estimating mutual information, the Voronoi diagram partitioning presents a challenge to defining coherent marginal distributions. This underscores the need to avoid relying on Voronoi diagrams and potentially even the Lebesgue measure for fast estimation, particularly in high-dimensional spaces where the performance of both concepts are degraded.  

We now present several approximation techniques to address these concerns that use different geometric measures and various means of partitioning the state space (with or without a known boundary) to successfully approximate the value given by the Voronoi-based definition, but in both deterministic and more efficient ways.

\subsection{Approximating Geometric Partition Entropy}
\label{subsec:Approx}
The first several methods of approximation rely on obtaining an initial grouping (or clustering) of the data set into $K$ near-equally represented geometrically distinct subsets.  As with the coarse-grained Voronoi estimate, some of the methods that follow can be applied to such clusterings obtained in any way. But, in order to utilize the geometries, we now require an associated partition of the underlying state space, $\mathscr{D}$, to avoid computing the Voronoi diagram.  

So, before presenting approximation methods, we first provide a deterministic algorithm and a simplification thereof, each of which not only provides an instance of the desired geometric-based clustering, but also defines a partition of $\mathscr{D}$ into rectangular boxes defining a set of $K$ macrostates, making several approximations straightforward to compute. 

\subsubsection{Factor Splitting Algorithms}
The Recursive Prime-Factor Splitting (RPFS) algorithm can be applied in either the standard $d$-dimensional Cartesian basis or subsequently after reorienting the data into a de-meaned Singular Value Decomposition (SVD) vector basis. If using the Cartesian basis, the process begins by first reordering the dimensions of the data set based on the marginal variances in each basis vector direction.  The prime factorization of the parameter $K$ is then computed, which we denote by $K=k_1\cdot k_2\cdots k_s$, where the $s$ factors, $k_i$, are ordered from largest to smallest. 

Beginning with the entire set $X$, at each stage in an iterative process of $s$ steps, the set of current subsets of the data are each split further into $k_i$ subsets along a common basis direction by computing a quantile-based partition of the subset of points projected onto the currently designated basis vector. These partitions use the $j/k_i$-th quantiles of the subsets of the data for $j=1,...,k_i-1$ together with the bounds of the current partition of $\mathscr{D}$ restricted to that region of the marginal space.  These quantile-based splittings are performed recursively on all current subsets while cycling through the $d$-dimensional basis vectors in the variance-based order.  

In order to avoid zero width quantiles during the various splitting procedures (e.g., dealing with points along a line or plane), the process can be treated with the same tolerance $\epsilon$ as before, but applied in the projected spaces, where if the number of unique projected values is smaller than the prime factor, that group can be split into a smaller number of bins or simply skipped, but the associated value of $K$ must then be adjusted after the process is completed to accurately reflect the size of the partition. 

Perhaps not surprising then, partitioning of this type works best when the parameter $K$ is taken to be $K=2^m$ with the integer $m$ being divisible by the dimensionality of the data, $d$. However, it can be applied for any choice of $K$, though the existence of large prime factors can lead to inaccuracies, which make estimations based on this form of clustering more sensitive to the choice in parameter $K$, as we will see in Subsec.~\ref{subsec:compare}.  Improvements in accuracy upon this basic approach are possible in terms of dynamically matching the largest remaining prime factors with the basis vector with the largest remaining variance, but for computational expediency, we instead focus on a further simplified version of this approach, which we call Factor Splitting.  

The Factor Splitting (FS) algorithm begins by first grouping the set of prime factors of the parameter $K$ into $d$ subsets whose products are either closest to being equal (or that best match the proportion of variances in an SVD basis).  We then apply the same quantile-based splitting approach as before, but restricting its application to a single factor in each dimension.  This simplifies the bookkeeping of boundaries but adds a minimal increase in the bias toward the first basis direction.  Due to the grouping of prime factors into larger dimension-based factors, the FS method will no longer necessarily benefit from the restriction to $K=2^m$, but similarly the quality of partition will improve with the equality (or alignment with the variances) of the $d$ factors.  Despite the potential variation with the parameter $K$, in our eventual application to Mutual Information estimation, we will choose $K$ in a structured way that largely negates these concerns.

The resulting groups obtained from either the RPFS of FS algorithms can then be used on their own for the Voronoi-based supercell estimate of GPE as described in the previous section, or in the next several approximation schemes.

\subsubsection{A Data-Driven SVD Estimator}
\begin{figure}[ht!]
    \centering
    \includegraphics[width=7.8cm]{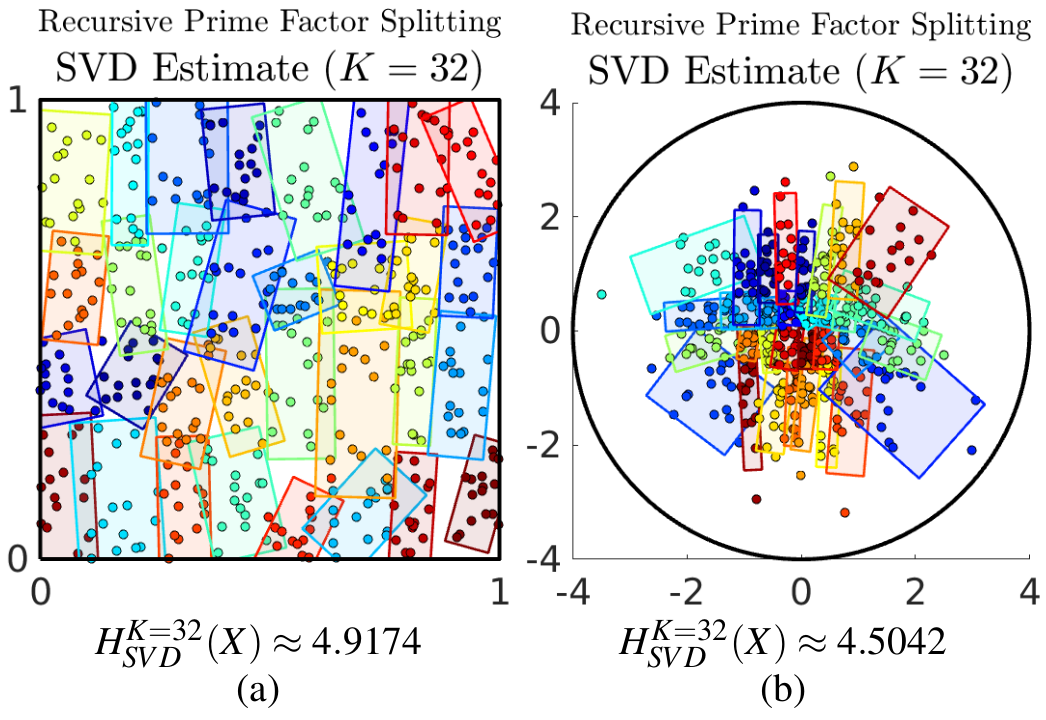}
    \caption{In the absence of known boundaries, the SVD of each subset of data can be used to produce a data driven geometric measure based on the product of singular values; and representative rectangular areas are shown for the RPFS defined partition for $K=32$ applied to the uniform and normal samples in (a) and (b) respectively, however, the normalization procedure makes these an abstract representation of the normalized measures.}
    \label{fig:svdapprox}
\end{figure}

In the absence of a known boundary, there is a simple data-driven estimator that can make use of any geometric clustering of the data.  Given $K$ clusters, the SVD of each subset of data is computed and the product of the singular values of each cluster is used as a geometric measure of the macrostates.  This measure is then divided by the frequency count of each macrostate as usual to form an estimate of local specific volumes, similar to $\mu$ in the previous section.  This approach generally performs well, but requires the computation of SVD subroutines in addition to some clustering routine prior to its application.  In addition, the data-driven bounds may not coincide with the true boundary of the region, which can result in large errors.  The SVD-defined areas are shown as rectangular regions corresponding to the RPFS defined clusters for the previous uniform and radially bounded normal samples in Fig.~\ref{fig:svdapprox} along with their respective entropy estimates.

It should be noted that the radial boundary shown in Fig.~\ref{fig:svdapprox} (b) is more spacious than the data-driven areas estimated by the SVD, and so this estimator will consistently overestimate the entropy of such a sample with large unobserved low probability regions of the state space.  This sort of discrepancy between data-driven bounds and true estimates will be a recurring theme for any region $\mathscr{D}$ with unobserved portions of the state space included since GPE estimates are designed to be sensitive to outlier impacts.

\subsubsection{The Box Estimator}
In addition to providing an equitable clustering of the data, both factor splitting algorithms also provide an associated partition of the state space into disjoint rectangular boxes.  If the boundary for $\mathscr{D}$ is known, these rectangular boxes should be intersected with $\mathscr{D}$ (or the SVD transformation thereof) before computing the geometric measures of the resulting macrostates.  Alternatively, if the data is believed to be generally dense in the domain (unlike the radially bounded normal sample), and $\mathscr{D}$ is mostly convex, the box estimate for GPE will not suffer from using the rectangular box measures directly (i.e. ignoring the boundary).  Similarly, if no boundary is known, the resulting boxes can be used as a form of data-driven boundary, though a similar concern to that referenced in the use of the convex hull arises here in that the boxes will more often include portions of $\mathbb{R}^d$ not in the true bounded state space $\mathscr{D}$, while also using extreme values from the data to define bounds of some boxes.  

\begin{figure}[ht!]
    \centering
    \includegraphics[width=8cm]{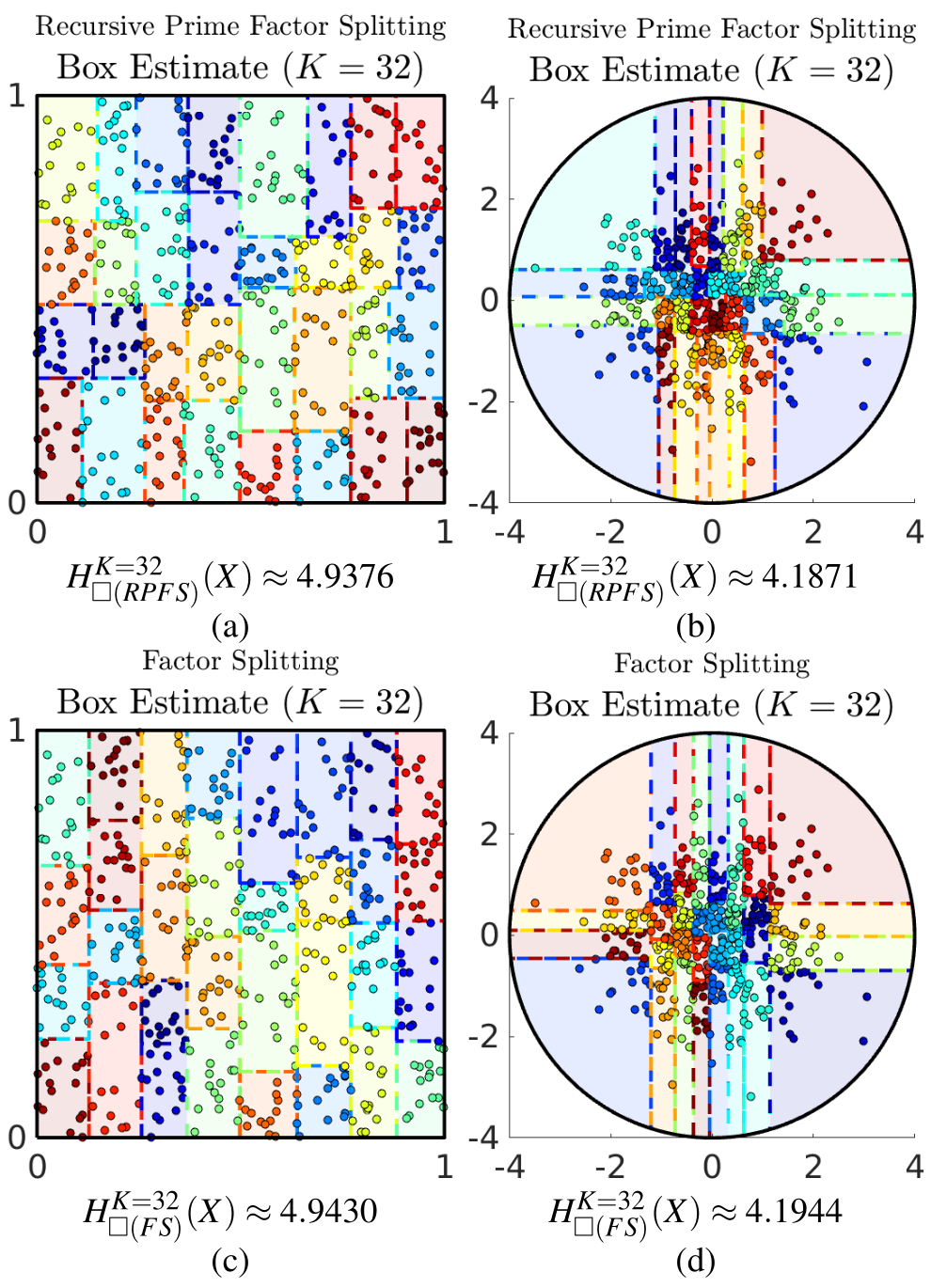}
    \caption{The RPFS-defined partition using parameter $K=32$ of the sample from (a) Figs.~\ref{fig:UniformVor} and (b)~\ref{fig:NormalVor}, where the boxes on the boundary are intersected with $\mathscr{D}$; along with the FS-defined partition for comparison in (c) and (d), respectively.}
    \label{fig:boxapprox}
\end{figure}

Whether the boundary is incorporated or not, using a measure based on Equ.~(\ref{equ:measure}), where the chosen geometric measure of the resulting boxes (potentially intersected with $\mathscr{D}$) is divided by the respective frequency count of the data set within each box, we can estimate the GPE using Equ.~(\ref{eqn:PE}).  As with the Voronoi definition, since the frequency count will be near uniform by definition (except in the presence of repeated values when projected in the marginal spaces), the measure is largely driven by the geometry.  It is thus important to realize that the approximations using the rectangular boxes directly (without intersection with $\mathscr{D}$) may exaggerate the geometry of some macrostates, especially when applied to skewed data.  So, if a boundary is known, it should be incorporated.  Figure~\ref{fig:boxapprox} (a) and (b) show examples of the RPFS-defined partition using $K=32$ for the same samples of uniform and multivariate normal distributions from Figs.~\ref{fig:UniformVor} and~\ref{fig:NormalVor}, respectively; whereas (c) and (d) do so for the simplified FS-defined partition for comparison.

One major hurdle remains in the generalization of GPE for broader applications in information theory. The clusters resulting from either the RPFS algorithm, the FS algorithm, or almost any other geometric clustering approach in a $d$-dimensional joint space still present the challenge of extracting marginal distributions.  If the marginal spaces of interest are the same dimensionality, we can adapt the above approaches, as will be done in Sec.~\ref{sec:MI} for the box (and Voronoi) estimator; however, the final approximation method for GPE was developed specifically to enable the development of a more versatile and efficient mutual information estimator, which will also be introduced in the subsequent section.  This approach as applied to GPE estimation for a general parameter value $K$ will tend to have large variation, and so it may not be ideal for this purpose.  However, its computational complexity and versatility are unmatched by any other estimator and so it is presented in order to provide a clear comparison.  

\subsubsection{The Partition Intersection Estimator}
\label{sec:pi}
Inspired by the original quantile-based approach in one dimension, we begin this estimation technique by computing a $k$-bin quantile-based partition of the entire data set projected onto each of the $d$ dimensions, resulting in separate partitions defined in the marginal spaces of each dimension.  These $d$ partitions are then intersected over the region in $\mathbb{R}^d$ to create a rectangular grid of $K=k^d$ bins of unequal size \textit{and} membership.  Note the designation of the lower case parameter $k$ here, which is used differently than $K$; alternatively, one can define a set of $d$ parameter values, i.e. $k_1,k_2,...,k_d$, whose product is equal to the $K$ used in other estimates for comparison, where the factors can be chosen based on the variation in each marginal space in some way.  

Since each of the partitions will range over the full set of observations, the resulting $d$-dimensional volume will necessarily contain $\mathscr{D}$, but will be rectangular in shape.  Further, since the intersecting partitions were defined in the marginal spaces, there is likely a wide range of frequencies of data observed within these $K=\prod_{i=1}^d{k_i}$ macrostates defined by the intersecting regions, including many that may have no observations at all.  Here we run into new problems using the measure~(\ref{equ:measure}), since $\mu(A_i)$ will be undefined whenever $\#_X(A_i)=0$.  We will consider an alternative measure later on, but for now we can set $\mu(A_i)=0$ for any such element of the $\pi$ partition, though this means that the measure theoretical explanation of GPE in terms of Kolmogorov and Sinai will no longer hold when the partition does not generate a $\sigma$-algebra on $\mathscr{D}$, and so here we are altering the domain to be some $\Tilde{\mathscr{D}}\subseteq\mathscr{D}$.  

The $\pi$ approach can still be applied in such cases though since, similar to histogram estimates, the entropy estimate becomes defined with respect to a new bounded region $\Tilde{\mathscr{D}}$ consisting of the union of all rectangular bins with non-zero measure.  Additionally, similar to the box estimator, the non-zero rectangular intersected partitions should be further intersected with the original boundary of $\mathscr{D}$ to ensure the exclusion of known regions that are not part of a valid state space.  As the parameter $k$, and therefore the number of bins used increases, the intersection with $\mathscr{D}$ becomes less important, but there is increased potential for removing unobserved space from $\Tilde{\mathscr{D}}$.  This impact on the domain must always be kept in mind when using the $\pi$ partitioning approach, and alternative measures can be used that utilized a term such as $\exp{(-\#_X(A_i))}$ in the measure to include unsampled bins as indicative of highly uncertain regions of the state space.

\begin{figure}[ht!]
    \centering
    \includegraphics[width=8cm]{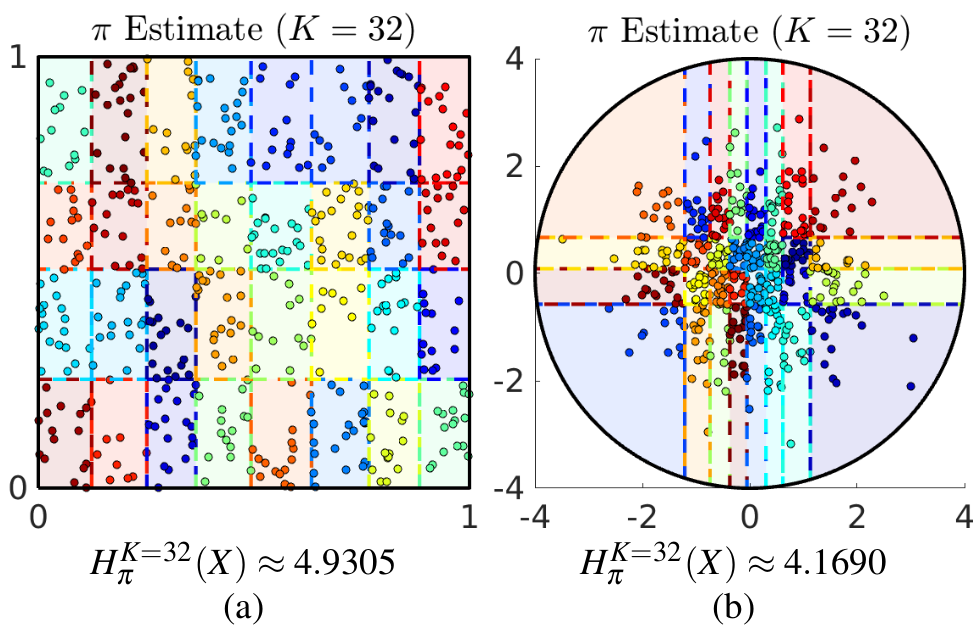}
    \caption{The $\pi$ partitions of the same samples of $N=500$ points drawn from (a) a uniform and (b) a multivariate normal distribution, resulting in good estimates as compared to the Voronoi-based estimates in Figs.~\ref{fig:UniformVor} and ~\ref{fig:NormalVor}, because of the symmetry and localization of the data.}
    \label{fig:PI}
\end{figure}

While it will often be beneficial to consider such alternative measures, we will initially use the specific volume measure defined in Equ.~(\ref{equ:measure}) with the added caveat of removing bins with no observations for the sake of direct comparison with the other methods. Figure~\ref{fig:PI} (a) and (b) show the partitions obtained from the Partition Intersection ($\pi$) estimate of GPE for the uniform and radially bounded normal samples above, respectively, where for consistency we have used $k_x=8$ and $k_y=4$ to produce $K=32$ bins.  Since both of these samples are generally localized in the region $\mathscr{D}$, we find relatively stable estimates for this measure, though the radially bounded sample from a normal distribution shows variation within the approximations due to how the boundary and related outliers get included.   

In fact, the measure defined in Equ.~(\ref{equ:measure}) can, in general, lead to an overemphasis on the importance of outliers in the $\pi$ partitioning approach, in particular, due to the variation in macrostate frequencies.  This is due the potential for low probability regions to lead to intersecting partitions that have a large geometric measure, which then get divided by the smallest frequencies, thereby increasing the overall importance of that state to the weighted entropy estimate.  This may be considered a desirable feature of this estimator in certain cases, and in fact, this characteristic of the $\pi$ estimator will feature prominently in our discussion of mutual information estimators when data includes informative outliers.  However, in the larger context, this can lead to a lack of stability in estimation with variation in the parameter $K$.  

\begin{figure}[ht!]
    \centering
\includegraphics[width=7.65cm]{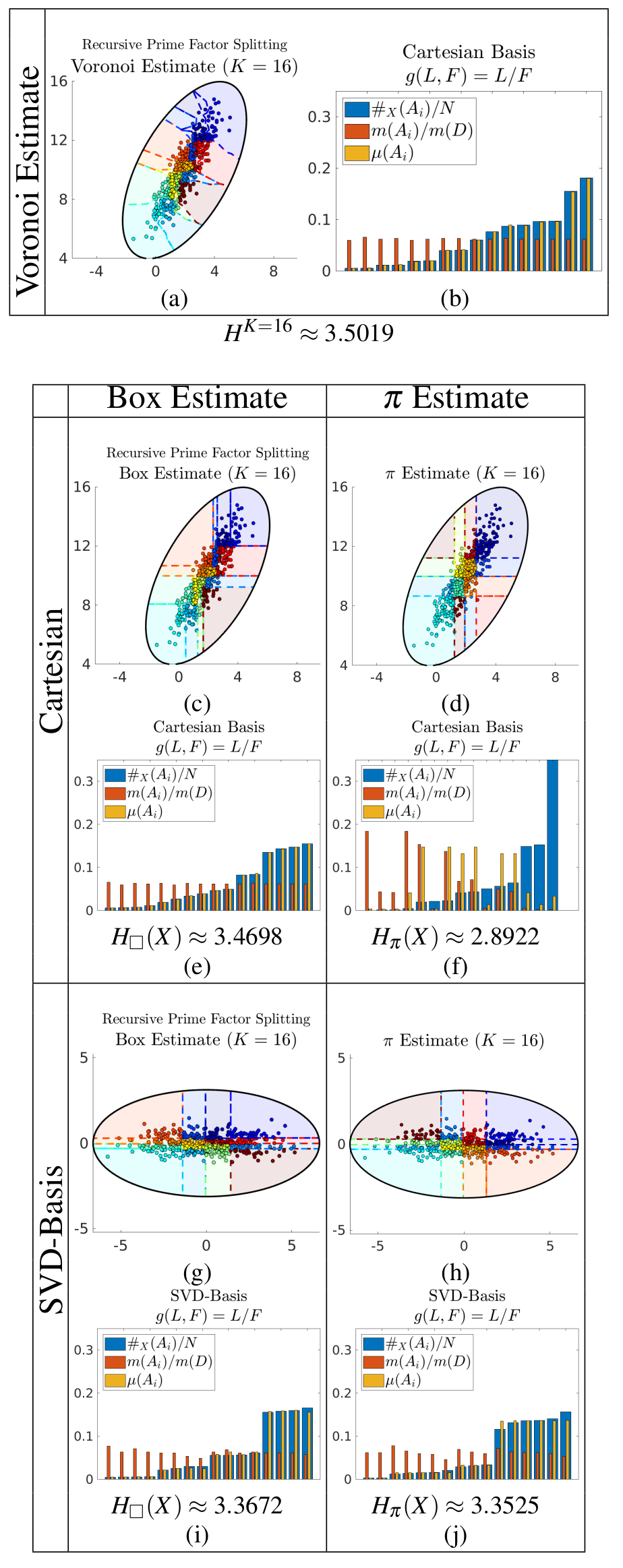}
\caption{(a) A skewed dataset in an elliptical boundary is partitioned by the RPFS algorithm, resulting in the Voronoi-based partition shown in (b); a comparison of the (c) RPFS-defined box partition and (d) $\pi$ partition illustrate the main failure of the $\pi$ estimator as seen in the associated sorted probability vectors (the blue bars) in (e) and (f) respectively; using an SVD basis as shown in (g) and (h) can improve performance shown in (i) and (j), where both estimates now approximate the Voronoi-based estimate more closely.}
    \label{fig:skewed}
\end{figure}

To illustrate this clearly, consider the skewed dataset shown in Fig.~\ref{fig:skewed} (a) consisting of a sample of $N=500$ points drawn from a multivariate normal distribution with correlation matrix $\Sigma=[~1~,~1.5~;~1.5~,~3~]$ bounded by an ellipse whose semimajor and semiminor axes are defined by the singular values of the data set, along with (b) a histogram showing the measure $\mu(A_i)$ in blue and each of the normalized component measures ($m(A_i)/m(\mathscr{D})$ as red and $\#_X(A_i)/N$ as orange) for the set of $K=16$ Voronoi-based supercells $A_i$ in a partition, which give the GPE estimate (for the RPFS-based clustering) of $H^{K=16}_{RPFS}(X)\approx 3.5019$. Fig.~\ref{fig:skewed} (c) and (d) show the resulting state space partitions defined by the RPFS and $\pi$ partitioning approaches, respectively, along with the corresponding histograms of their subsequent measures in (e) and (f).  While the RPFS partition does a good job estimating the Voronoi-based value, the $\pi$ partitioning fails due to the large deviations in the measure $\#_X(A_i)$ for the resulting partition.  In this instance, some of the macrostates that are counted as the most important actually have very little geometric measure, but even less relative frequency, meaning it is arguably the least relevant states that are weighted most in this case. 

The performance of the $\pi$ estimator can be improved through a coordinate transformation into an SVD-defined basis, resulting in the comparable plots in Fig.~\ref{fig:skewed} (g)-(j), which give GPE estimates of $H_{\square}^{K=32}(X)\approx 3.3672$ and $H_{\pi}^{K=32}(X)\approx 3.3525$; which are much better approximations of the more accurate Voronoi-based estimate, but this will not always succeed in improving the estimate.

Thus, the $\pi$ estimator is expected to suffer whenever the distribution of frequencies is large, and this is especially so in the case where there are any regions with very small, but nonzero frequencies as compared to the average.  For this reason, the $\pi$ estimation of GPE with the standard choice of $\mu$ is only suggested for data sets having minimal outliers or for choices of $k$ that result in some minimum number of observations in each non-zero membership box, as even a single outlier can alter the estimate drastically when the distribution is normalized.  Additionally, this challenge can be addressed by first defining the states and then adjusting them based on nearby frequency counts to improve performance, but this would be a time intensive non-automated task.

The general sensitivity to outliers will be harnessed in the following section to enable a dynamic class of new mutual information estimators that can be tailored for various analysis tasks such as those involving the transient dynamics in synchronization.  However, we now suggest an alternative set of measures that can be useful when coupled with any of the estimators, but are particularly helpful for regularization of the $\pi$ estimator on data sets with low probability regions.

\subsection{Alternate Measures for the $\pi$ Estimator}
In practice, if outlier impacts are considered less important to the dynamics in question or one is specifically seeking results that coincide with more traditional methods, meaning the averaged mutual information is sought, which essentially ignores outliers, then it is also possible to de-emphasize the impact of outliers by using an alternative measure.  If we consider a vector of $M$ geometric measures of the macrostates, denoted by $L$, and a vector of $M$ macrostate frequencies (including repeated microstate observations), denoted by $F$, then we can define a measure using any functional form of these two quantities, $g(L,F)$, after appropriate normalization.  Our local specific volume measure $\mu(A)=\frac{m(A)\#_X(A)}{N m(\mathscr{D})}$ is then associated with the functional form $g(L,F)=L/F$.  This choice indicates that we place importance on $L$ for the quantification of uncertainty while considering higher values of $F$ in a macrostate to indicate less uncertainty (since it has been sampled more).  When applied to equitably represented macrostates, this performs much like the original GPE. However, since this choice divides by $F$, it is sensitive to any smaller values in $F$.  While this is not generally a problem in any estimate except the $\pi$ estimator, it is also possible for large numbers of repeated values to affect the accuracy of this choice in other estimators.  

In contrast, it has been observed that using a measure of the functional form $g(L,F)=L\cdot F$ provides a good balanced quantification that produces results that generally coincide with histogram-based methods due to a reduction in outlier importance in the $\pi$ estimator.  Estimates computed using the $\pi$ partition and this measure may be thought of as a weighted entropy measure that is specifically designed to reduce the impact of outliers, since low frequency regions will be multiple times smaller than those more heavily populated.  And while other measures will incorporate the impact of outliers in different ways, interestingly, this correspondence further indicates the potentially problematic treatment of outliers by basic histograms.  

In addition to these two main choices, we will also consider two other functional forms when the variation of both $L$ and $F$ are large.  The functional form $g(L,F)=F\cdot e^{-L}$ allows for partitions with highly repeated values (e.g., periodic dynamics) to incorporate the geometry in a meaningful way while bounding its effect on the mostly categorical data represented by highly repeated values; while the use of the functional form $g(L,F)=L\cdot e^{-F}$, alluded to in the previous section, enables the proper incorporation of the domain with the $\pi$ estimator even when some elements of the partition have little to no representation.

\subsection{Comparison of Estimators}
\label{subsec:compare}
\begin{figure}[ht!]
    \centering
\includegraphics[width=8cm]{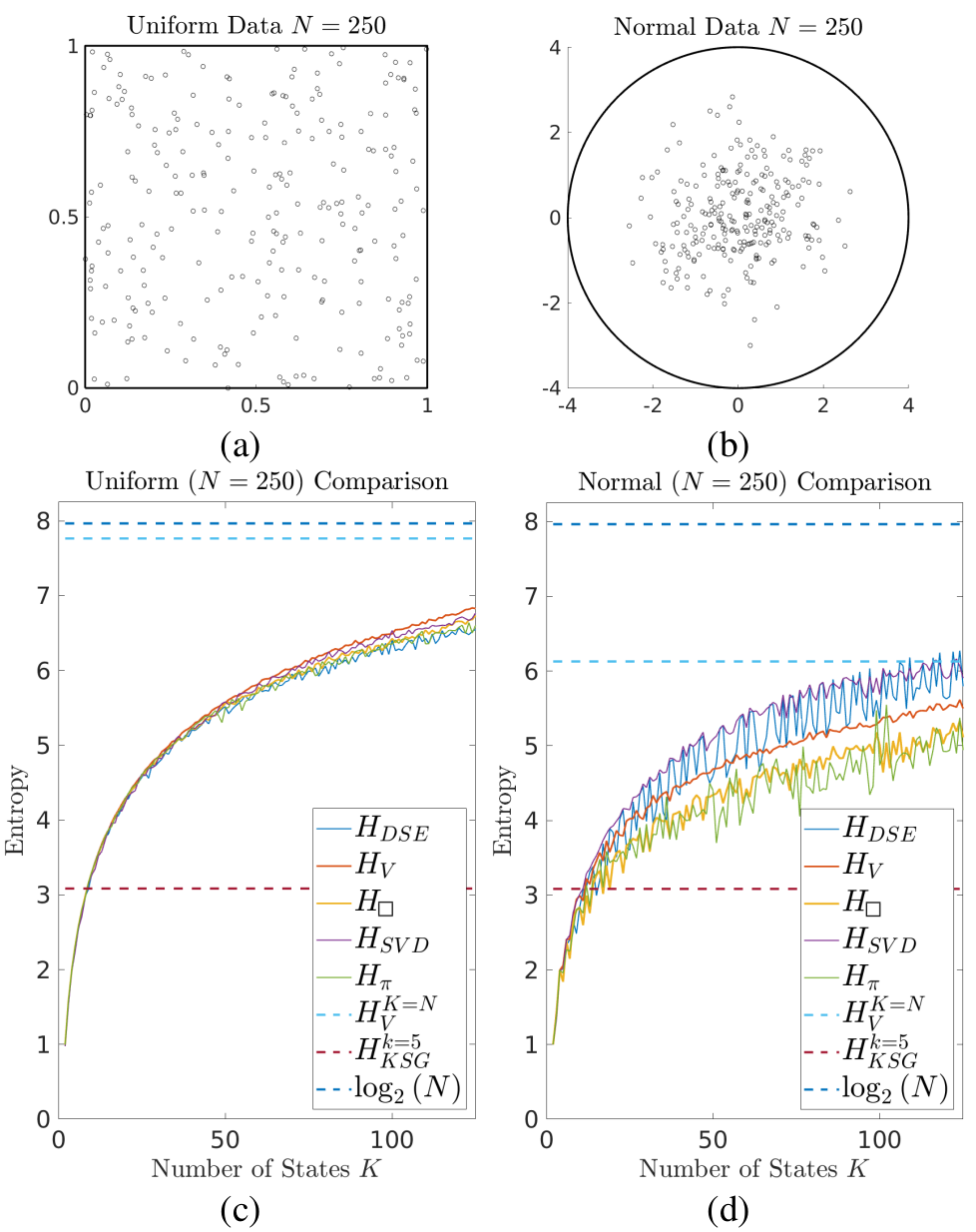}
\caption{Comparison of the various entropy estimators over a range of the parameter $K$, including four estimates of GPE along with a 2D histogram-based estimate of DSE (where the number of bins in the marginal spaces are defined by the prime factorization and used in all estimators for consistency); the KSG estimate $I(X,X)$ is provided for reference for the choice of $k=5$ nearest neighbors along with some of the limiting values as dashed lines for comparison.}
    \label{fig:compareGPE}
\end{figure}

As in the one dimensional setting presented in~\cite{diggans2022geometric}, the limiting values of GPE and DSE estimates as $K\rightarrow N$ and $N\rightarrow \infty$ will only converge to the same asymptotic value for uniform distributions, converging in the limit as $\log_2{(N)}$.  For the uniform sample of size $N=250$ shown in Figure~\ref{fig:compareGPE} (a), this convergence is shown (within sampling variability) in (c), where the basic histogram estimate of DSE is compared with four estimators of GPE: i) the RPFS-clustered Voronoi estimate, ii) the RPFS-defined box estimate, iii) the RPFS-clustered SVD estimate, and iv) the $\pi$ estimate, all using the measure defined in Equ.~(\ref{equ:measure}) of the form $g(L,F)=L/F$. The constant values of $\log_2{(N)}$, the $N$ cell Voronoi estimate, $H_V^{K=N}(X)$, and a KSG estimate, $H_{KSG}^{k=5}(X)=I(X;X)$ are included as dashed lines for reference.  The estimates are plotted over the parameter range $K\in[2,125]$ since larger values lead to degenerate volumes for the SVD approximation.  

Furthermore, we have used the component parameter factors $k_x$ and $k_y$ (where $K=k_x\cdot k_y$) resulting from the RPFS algorithm for all other estimators, including the DSE estimate, in order to obtain as consistent as possible comparisons.  It is clear from this plot that all of the GPE estimates are basically consistent with the histogram DSE estimate for uniform samples.  The Voronoi-based GPE estimator is the most stable with respect to changes in the parameter $K$, but all GPE estimates follow a similar trend line as $K\rightarrow N$ approaching the unique $K=N$ Voronoi estimate (which is slightly less than $\log_2{(N)}$ due to sampling variability). 

The same generally holds for the sample from the radially bounded normal distribution, though we see more variability in the estimates with changes in the parameter $K$.  Again, we see the SVD approach overestimate due to not incorporating the true boundary, while both the box and $\pi$ estimators are underestimating the GPE when compared with the Voronoi estimate due to a more uniformly distributed geometric area resulting from the box partitions.  The DSE estimate eventually increases over the GPE estimates due to its eventual bias toward $\log_2{(N)}$, indicating the potential benefit of GPE estimates.

\begin{figure}[ht!]
    \centering
\includegraphics[width=8.3cm]{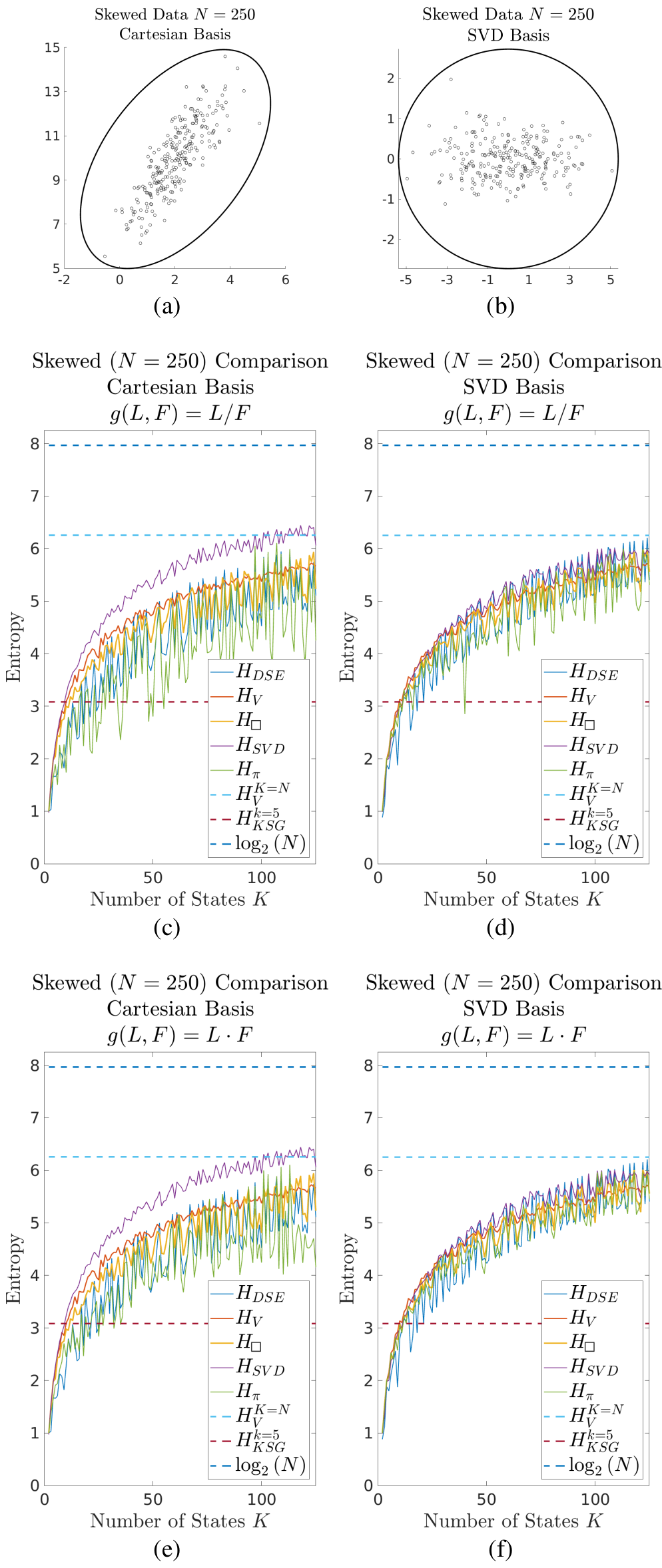}
\caption{The same comparisons as in the previous figure shown for the dataset shown in (a) along with its transformation into an SVD-bases coordinate system (b); (c) and (d) show the comparisons using the measure $g(L,F)=L/F$, whereas (e) and (f) show the comparisons using the measure $g(L,F)=L\cdot F$, where in (f) the $\pi$ estimator becomes as stable as the DSE estimate with respect to changes in the parameter.}
    \label{fig:compareskewedGPE}
\end{figure}

We now consider a skewed data set in Fig.~\ref{fig:compareskewedGPE} (a) consisting of a sample of $N=250$ drawn from the same correlated multivariate normal distribution as before. Although the DSE estimate in (c) is now reduced due to the larger regions outside the boundary that get removed along with several bins within the region $\mathscr{D}$ with no representation; this leads to an effectively lower $K$ value for the bias of the DSE estimate.  The general trends from the basic radially bounded normal sample hold for the comparisons of the GPE estimators in (c), however, as expected, we see a much larger instability in the values from the $\pi$ estimator.  These fluctuations can be reduced in several ways: i) we can convert the data set into an SVD-based coordinate system (as shown in (b)) prior to entropy estimation, which leads to the trends shown in Fig.~\ref{fig:compareskewedGPE} (d), but alternatively, we could also use a different measure such as the one corresponding to $g(L,F)=L\cdot F$, which results in the plots shown in (e) and (f) in the original basis and the SVD basis, respectively.

\begin{figure}[ht!]
    \centering
\includegraphics[width=7.5cm]{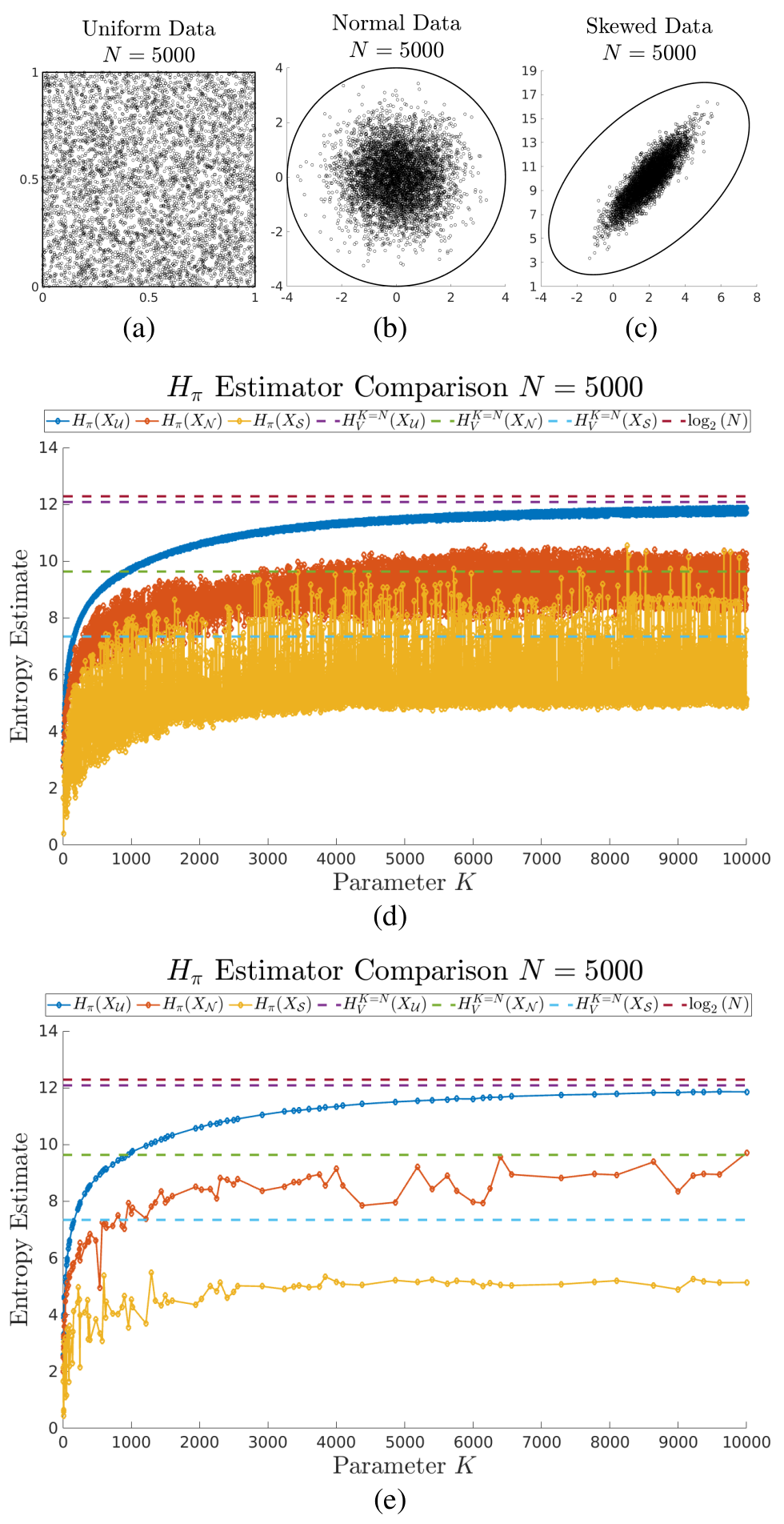}
\caption{Samples of size $N=5000$ from the prior uniform, normal, and skewed distributions are shown in (a), (b), and (c) respectively, together with (d) the $H_\pi$ estimation for all non-prime parameter values in the range $K\in[0,10000]$ and (e) the $H_\pi$ estimation for all parameter values in the range $K\in[0,10000]$ with only even number of prime factors with $k_i\leq 5$ for all $i$.}
    \label{fig:PIcompare}
\end{figure}

Since we will focus on the use of the $\pi$ partitioning approach for estimating mutual information in the next section, we wish to explore the variability of its estimates further.  Figure~\ref{fig:PIcompare} compares just the $\pi$ estimator applied to larger samples of $N=5000$ points drawn from the three distributions used thus far shown in (a), (b), and (c), where the boundaries have been adapted to ensure all points are in the bounded region $\mathscr{D}$.  Since the $\pi$ estimator is based on the computation of quantiles of the entire data set in each of the marginal spaces, this allows the use of many more macrostates than there are data points.  Of course, this will generally saturate since only those bins that contain any observations will be used in the calculation of entropy (unless $g(L,F)=L\cdot e^{-F}$ is used).  

Furthermore, the estimates are almost always underestimating the Voronoi-based estimate for $K=N$.  The estimations stay relatively consistent (although noisy) as the number of bins increases, but the large variability is more a feature of splitting of each $K$ value into the two factors $k_x$ and $k_y$, such that $K=k_x\cdot k_y$.  When $K$ is prime or has large prime factors, this process can result in very poorly distributed partitions, which leads to the larger variations.  Despite this draw back, we do not find a large bias toward the number of bins used, as long as we use a large enough $K$, e.g., $K\approx N$, though a bias toward the sample size will remain.  

In panel (e), we see that by restricting estimates to only those parameter values having an even number of prime factors and $k_i\leq 5$ for all $i$, a much smoother variation in the $\pi$ approximation results.  Some of the variation remains depending on the sample and distribution of outliers, but this restriction largely limits the variability, indicating that the $\pi$ estimator can be reliable if the number of states is divisible into $d$ roughly equal factors, thus we can suggest using any parameter of the form $K=k^d$ for some integer $k$.

Although all examples presented so far have been two dimensional for ease of presentation and comparison, higher dimensional applications will be explored through the lens of mutual information, which we now turn to as the main contribution of this paper.

\section{Geometric Partitioning for Mutual Information Estimation}
\label{sec:MI}
Many of the most intriguing entropy-based concepts in the study of complex systems (e.g., transfer entropy) are defined in terms of mutual information.  So, for this new viewpoint on entropy estimation to be more widely applicable to complex systems research, the concepts of GPE must be leveraged into appropriate mutual information estimators.  Here, we provide two computationally efficient approximation schemes for mutual information based on the box and $\pi$ estimators of GPE, which enable the once limited tools of high dimensional information theory to now apply in many new contexts.  In particular, the ability to choose various measures within each method makes them versatile and provides tailored approaches to mutual information estimation for research in complex systems where outliers may be informative for understanding the dynamics.

\subsection{The Partition Intersection Estimator of MI}
The partition intersection ($\pi$) approach that was used to partition the state space $\mathscr{D}$ for GPE estimation in subsec.~\ref{sec:pi} began by first defining quantile-based partitions in the marginal space of each dimension. These one dimensional partitions were then intersected within the overall state space, resulting in a rectangular non-uniform grid of boxes that cover the state space $\mathscr{D}$.  By design, this enables a straightforward definition of a new mutual information estimator that remains grounded in the basic concept of GPE.  However, as before, for the ideas of Kolmogorov and Sinai to apply, we must consider the data set on an alternate bounded region $\Tilde{\mathscr{D}}$, consisting of only those bins with nonzero frequency of observation, potentially intersected with the original set $\mathscr{D}$ to remove regions that are known to be outside of the true state space.  This means that low probablity regions and unobserved portions of the theoretically known state space might be removed, affecting the estimate, similar to what happens with DSE in high dimensions, though much less due to the data-driven bin definitions. Again, this can also be altered by the choice of measure, e.g., $g(L,F)=L\cdot e^{-F}$.

The $\pi$ estimator for MI, which we denote by $I_\pi(X;Y)$, simply uses the $\pi$ partitioning approach for estimating the non-conditional entropies in the joint and marginal spaces, then making use of Equ.~(\ref{equ:MI}). If $X$ and $Y$ are one dimensional, then the estimates for $H_\pi(X)$ and $H_\pi(Y)$ are simply the quantile-based estimates of GPE from~\cite{diggans2022geometric} with the newly added weighting of macrostate membership (including repeated values).  However, if $X$ and $Y$ reside in higher dimensions, their entropies are computed very much in the same way as the joint entropy, except restricted to the lower dimensional marginal subspaces.  

This approach requires the computation of quantile-based partitions in each dimension's marginal space, followed by binning the data within the unequal grid partition that is defined by the intersecting quantiles, where the product of the differences in nearby quantiles provides the geometric measure. In light of the large number of algorithms for obtaining quantile estimates~\cite{chen2020survey}, the $\pi$ estimation of MI provides an optimal approach with respect to computational complexity, while also being adaptable to any context, e.g., $d_1\neq d_2$.  The added benefit of the flexibility in the choice of measure (not only restricted to this estimator) also dynamically enables the inclusion of outlier impacts across applications.  

For example, as before, the use of the alternative multiplicative measure of the form $g(L,F)=L\cdot F$ obtains similar results for MI as the histogram-based DSE and is as consistent as the KSG estimator, supporting the claim that both of these traditional estimators de-emphasize the impact of outliers.  Other functions of $L$ and $F$ may also be explored for varying the weighting of outlier impacts, but we restrict our analysis to the cases described previously for brevity.

\subsection{The Box and Voronoi Estimators of MI}
Although none of the other partitioning algorithms from Sec.~\ref{sec:GPE} enable clearly defined marginal distributions for general applications where $d_1\neq d_2$, several other MI estimators are possible in the unique case when $d_1=d_2$.  Whenever the data sets $X$ and $Y$ occur in the same underlying state space $\mathscr{D}\subset\mathbb{R}^d$, i.e. $d_1=d_2=d$, then the joint space $\mathscr{D}\times\mathscr{D}\subset\mathbb{R}^{2d}$ is the product of two identical copies of the marginal state space.  By concatenating the two data sets $X$ and $Y$ in series within the shared space $\mathscr{D}$ (rather than in parallel in the joint space), a single partition of $\mathscr{D}\subset\mathbb{R}^d$ can be computed using either of the factor splitting algorithms (RPFS or FS) on the combined data set.  Similarly, the Voronoi diagram of the concatenated data set can be constructed to obtain $2M$ (or fewer after further adjusting for repetition across $X$ and $Y$) cells within the single copy of the state space. Subsequent geometric clustering, such as CKM, can lead to a shared partition of the marginal space $\mathscr{D}$, however we chose to focus on the box approximation due to its scaling in higher dimensions.

It is important to note that the macrostates defined in this way will now not necessarily have nearly equal numbers of points from both $X$ and $Y$, but rather an equal number of points from the combined data set.  Thus, it is reasonable to expect similar challenges to those found with the $\pi$ estimator in that large variances in frequency counts from $X$ or $Y$ can lead to poor estimates.  For this reason, which is especially true in higher dimensions, it may be best to use a relatively small parameter value, e.g., $k=3,4,5,$ or $6$ in each dimension; and, although a bipartition in each dimension ($k=2$) may struggle to capture detail in lower dimensional applications, it performs surprisingly well as the dimension increases.  Further, this approach scales well with comparison across dimensions, since the overall number of bins used will scale exponentially with dimension, i.e. $K=k^d$.  For our purposes, we will generally use $k=6$, except for low dimensional examples, e.g., $d_1=d_2=1$, where finer grained detail will be useful.
For example, when analyizing synchronization of one dimensional maps in Sec.~\ref{sec:results}, we use $k=24$ to match the partitioning of the $\pi$ estimator for comparison.

Since the computation of the Voronoi diagram and even the RPFS-partitioning procedure become more intensive in higher dimensions, we restrict our consideration to the simpler Factor Splitting (FS) approach to define a box estimator for MI. Using the partition obtained from the FS algorithm applied to the combined data $(X;Y)\subset \mathbb{R}^d$ (in series), the joint entropy $H(X,Y)$ is estimated using the symbolized joint states in $(X,Y)$ where the products of the geometries in $\mathbb{R}^d$ for each element of the corresponding symbols in $X$ and $Y$ are divided by the joint symbol frequencies.  The marginal entropies $H(X)$ and $H(Y)$ are then computed using the box estimator with the same FS-defined partition of $\mathscr{D}$.  Using these component entropy estimates, the box estimator of mutual information follows from Equ.~(\ref{equ:MI}). 

One final point to make on this approach is that if the mutual information is high and or a small enough parameter $k$ is chosen, the FS partitioning procedure will be more likely to include large portions of low probability or even unobserved state space when compared with the $\pi$ partitioning approach (unless using the $g(L,F)=L\cdot e^{-F}$ form of measure). This is the main difference between the two partition-based MI estimators, and this feature will be explored in the next section, but care should be taken in deciding what approach is more appropriate for a given data set or state space context.  

\subsection{Dealing with Negative Estimates of Mutual Information}
\label{subsec:negMI}
By definition, the Mutual Information should always be non-negative.  Many MI estimators, but especially the KSG and other knn-based approaches, still give negative values often due to the variation in biases of the marginal and joint entropy estimates.  When all observations are considered to be of equivalent informational value, frequency-based measures are biased in predictable ways, and so in general, these negative outcomes are often small in size, and it has been common practice to simply set negative values equal to zero when they arise.  However, when leveraging the geometry in a way that exaggerates the role of outliers, those differences in bias between marginal and joint entropy estimates can become much larger.  Thus, we find that both the $I_\pi$ and $I_\square$ estimators can take on much larger magnitude negative values of MI, which is problematic.  We therefore must understand when these challenges arise and provide an adaptive solution.  There are two cases to consider: the first is where the outlier impact becomes exaggerated in the joint space due to the curse of dimensionality; and the second arises in the presence of highly repeated values, where the data may be better considered as categorical rather than truly on a continuum. 
In the second, detailed analysis on the change in entropy with increased tolerance $\epsilon$ can be informative here.  But, regardless of its cause, whenever negative estimates are encountered in either the $\pi$ or box estimators, instead of artificially setting those values to zero, we dynamically alter the measure $\mu$ to be of the form $g(L,F)=F\cdot e^{-L}$.  Recall that the frequency vector $F$ is for a set of (unequal measure) geometrically defined bins, meaning the informational value of each observation is proportional to the size of the bin in which it is found, with larger bins implying more informative observations.  The multiplication by $e^{-L}$ in this measure then reduces the importance of the geometry slightly.  This reduction of importance placed on geometry coupled with the increased attention to frequency (though still defined on non-equal measure partitions) results in more meaningful estimates in the case of highly repeated values, while also reducing the impact of large geometries in the joint space. 

For instance, when exploring applications to synchronization in the next section and the dynamics result in periodic orbits (or otherwise produce negative MI estimates), the MI recomputed using a measure of the form $g(L,F)=F\cdot e^{-L}$ is found to result in a smooth transition from estimates using $g(L,F)=L/F$, while remaining positive, in such cases.

\section{Results}
\label{sec:results}
We begin with an analysis of the computational complexity of the two partition-based MI estimators, comparing their performance with both the histogram-based DSE and KSG estimators of MI.  This is followed by a study of how these new MI estimators' performance scales with increased dimensionality, where a comparison with the known ground truth is possible for a simple experiment that was designed to explore highly dependent Gaussian random variables.  We then provide some interesting applications in synchronization dynamics where our approach enables an entirely new form of analysis on transient dynamics.

\subsection{Computational Complexity}
Although time complexity is often featured in discussions of computational complexity, both time and space complexity are important metrics in determining the efficiency and practicality of an algorithm.  An algorithm that runs quickly but consumes excessive memory can be impractical for large-scale applications, and this is specifically a problem with the KSG estimator.  

Furthermore, the complexity of nested algorithms that call many subroutines, such as mutual information estimators, is challenging to accurately portray in standard notation as it will often rely heavily on the choices made between various implementations within the overall method.  Often times there are trade-offs for approximations and speed versus storage.  For instance, the KSG estimator's time complexity is driven by the identification of nearest neighbors, which achieves its best time complexity using the KDTree algorithm with a time complexity of $\mathcal{O}(N\log{}N)$ for the average case and $\mathcal{O}(N^2)$ for the worst case. However, the storage complexity of the KDTree routine is $\mathcal{O}(N)$ \cite{chen2020survey}.  This limits its use to sample sizes on the order of tens of thousands for most common hardware, which would be fine, except that mutual information in particular can require even larger sample sizes for high dimensional data with large dependencies.

On the other hand, recent advances in quantile computation reached a time complexity of $\mathcal{O}(N)$, and impressive storage complexity of $\mathcal{O}(\log{}\log{}N+ \log{}\frac{1}{\epsilon})$ for some algorithms \cite{chen2020survey}, where $\epsilon$ is the approximation error. Specifically, the heavy reliance on the computation of quantiles in both the $\pi$ and box estimators lead to a wide range of complexities that stem from these choices.

In addition, when comparing algorithm performance empirically, several factors must be considered beyond just time and space complexity. The quality of the implementation can vary, with one algorithm potentially being better written than the other, leading to biased results, and the choice of test cases can also impact the comparison, potentially favoring one algorithm unfairly. Moreover, the processor and processing methodology (parallel, sequential) can significantly impact the results of comparisons as well. 

Thus, direct comparisons are provided between performances rather than discussing the theoretical complexity of the methods developed.  In an effort to de-bias the results, we used a vectorized, highly efficient implementation of the KSG estimator that showed the best performance in \cite{almomani2020entropic, fish2021entropic, almomani2020erfit}.  We perform all experiments on the same machine, a Titon 64-core server, running samples for fixed parameter choices in parallel, and using common built-in functions of the chosen language (Matlab) for implementation, e.g., $quantile(\cdot)$.

\begin{figure}
    \centering
   \includegraphics[width=8cm]{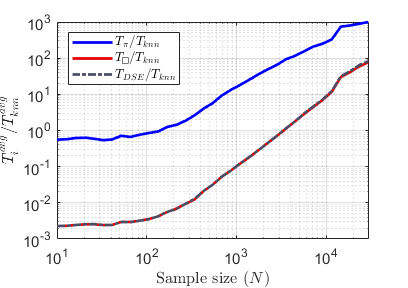}\\
    (a)\\
   \includegraphics[width=8cm]{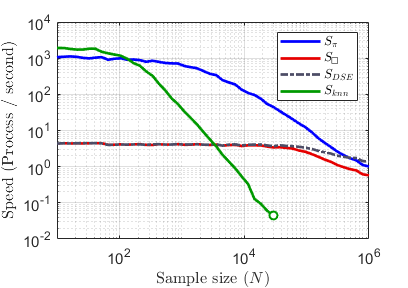}\\
   (b)\\
   \caption{Time complexity: the speed of each estimator (executions or processes per second) computed as $S_i = (Time_i^{avg})^{-1}$ is plotted in panel (a), while panel (b) shows the ratios of the average time to process the $\pi$, box, and DSE estimators over the time required for the KSG (knn) estimator on log-log axes.}
    \label{fig:Time}
\end{figure}

In Fig.~\ref{fig:Time} (a), we see the estimator speed (number of calls, executions, or processes that can be completed per second), computed as the inverse of the time averaged over 100 runs to avoid anomalous results from randomly generated data sets.  We see that the KSG estimator speed drops exponentially to be less than 1 process/second by the time $N=10,000$. This time requirement of the KSG method highly limits its application when dealing with complex networks or a large number of time series, where there is a need to perform mutual information estimations numerous times. Moreover, its space complexity prevented its continued use when the sample size reached $30,000$ due to insufficient memory on the available hardware; the circle indicates the last data point that was possible with the KSG estimator on the hardware used. 

In Fig.~\ref{fig:Time} (b), We plot the ratios of the time to compute each estimate using the efficient implementation of the KSG estimator as the baseline, since it is generally considered the state of the art in MI estimation.  The KSG method proves to be faster for small samples $N<100$ (ratio less than 1) due to the minimal computational overhead cost, and so we plot these ratios on a log-log scale for increasing sample size.

Here we see that once the sample size is large, i.e. $N>30,000$, the $\pi$ estimator becomes at least $1000$ times more efficient than the KSG estimator and $10$ times more efficient than the histogram-based and box estimators.  And, while the box estimator is less efficient due to the FS partitioning procedure, it remains on par in empirical time complexity with the histogram estimator, as indicated by their curves overlapping almost entirely.

\subsection{High-Dimensional Random Variables}
Considering how these estimators scale with dimension, we provide a simple experiment in the rare case of a known ground truth. For each choice in dimension, $d$, we generate two samples \(X\) and \(Y\) of size $N$, where the \(i\)-th dimensions, \(x_i\) and \(y_i\), are functions of the same random variable \(t_i \sim \mathcal{N}(0,1)\). Specifically, we define \(x_i(t_i)\) and \(y_i(t_i)\) as cubic polynomial functions of \(t_i\) as follows

\begin{eqnarray}
    x_i(t_i) & = & a_0 +a_1t_i +a_2t_i^2 + a_3t_i^3 \notag \\
    y_i(t_i) & = & b_0 +b_1t_i +b_2t_i^2 + b_3t_i^3
\end{eqnarray}
where the coefficient vectors \(\mathbf{a}\), \(\mathbf{b}\) \( \sim \mathcal{N}(0,1)\) are generated randomly in each iteration. This procedure results in high mutual information that scales linearly with the dimension.

Considering Gaussian i.i.d. random variables, the parametric estimator of mutual information can be defined as:
\begin{equation}\label{equ:cov}
I(X ; Y)=\frac{1}{2} \log \left(\frac{\operatorname{det}\left(\Sigma_X\right) \operatorname{det}\left(\Sigma_Y\right)}{\operatorname{det}(\Sigma)}\right)
\end{equation}
where $\boldsymbol{\Sigma}$ represents a covariance matrix for the random variables whose $(i,j)$-th element is given by:
$$
\Sigma_{i j} \equiv \overline{x_i x_j}-\left(\overline{x_i}\right)\left(\overline{x_j}\right).
$$
The subscripted covariance matrices in Equ.~(\ref{equ:cov}) are those defined in the appropriate marginal distributions, i.e. for $X$ and $Y$, while the unsubscripted matrix represents the full $2 d \times 2 d$ dimensional covariance matrix.

\begin{figure}[ht!]
    \centering
    \includegraphics[width=7.5cm]{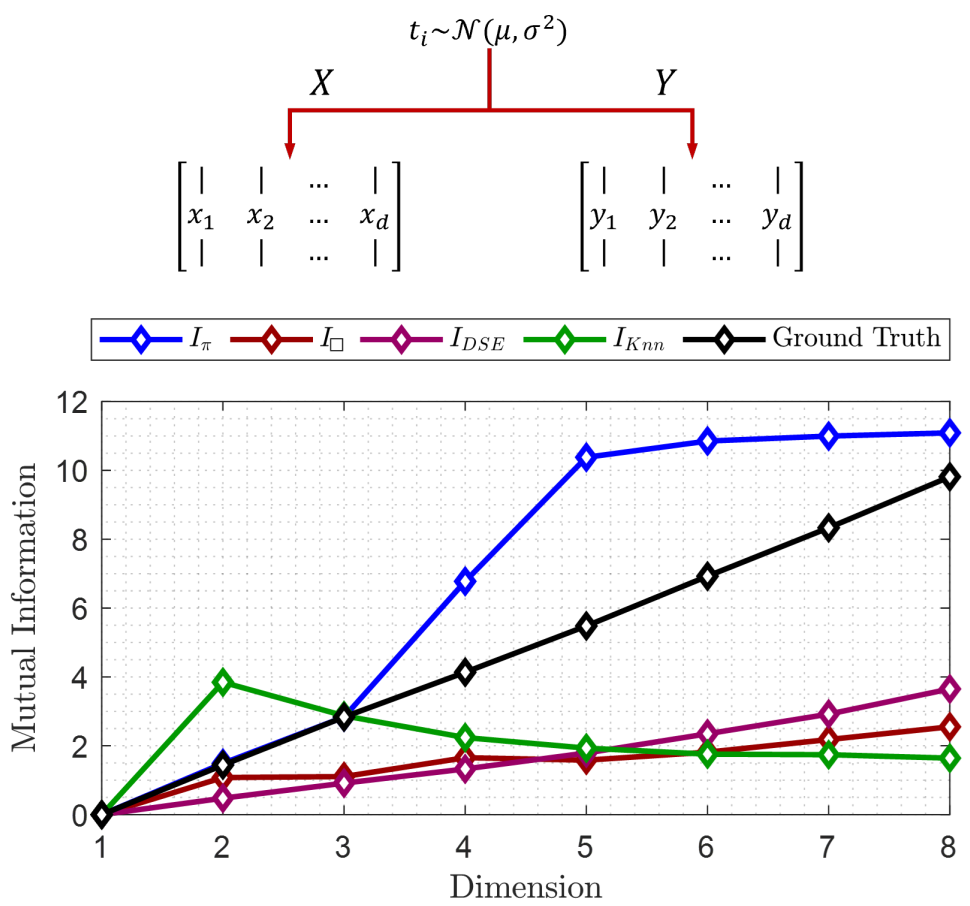}
    \caption{For each dimension, $d$, the MI estimates are averaged over $100$ runs and compared with a known ground truth computed using Equ.~(\ref{equ:cov}), where samples $X$ and $Y$ are generated by random coefficient cubic polynomials evaluated on the same normally distributed sample of the independent variable $t_i$.}
    \label{fig:Ex01_RandomSample}
\end{figure}

Here, we see that the true mutual information between the samples increases linearly as the dimension increases.  This illustrates one of the significant failures of the KSG estimator, seen in the decreasing estimate of MI. As the number of dimensions increases, the volume of the space increases exponentially, and so data points become sparser, making it difficult to find the true nearest neighbors. The increased sparsity causes the distances between points to become more uniform, reducing the effectiveness of the locally-focused KSG approach. Moreover, the volume of a neighborhood defined by the $k$ nearest neighbors increases with dimensionality, and as this volume grows, it tends to encompass more points, leading to an overestimation of densities and an underestimation of mutual information (due to $k$ and $N$ remaining constant in Eq.~\ref{eq:ksg}).

Although the histogram-based DSE estimate and the box estimator capture the general linear trend, they both struggle to scale accordingly due to the growing sparsity of the data in high dimensions. This sparsity is indicative of the curse of dimensionality, and so basically none of the non-parametric methods will be able to overcome this challenge entirely. In contrast, the $\pi$ estimator shows superior performance, however, this is believed to be due to the large empty spaces in high dimensions being ignored by the $\pi$ partitioning procedure, while retaining relevant information on the data that is present. This adaptive restriction of the domain to some smaller region $\Tilde{\mathscr{D}}$ can be seen as a shortcoming in other contexts, but it did give an accurate estimation of the true mutual information up to three dimensions and captured the general trend of increasing mutual information for higher dimensions as well. The KSG and DSE estimators stand on the assumption that the local density within each knn neighborhood or bin, respectively, is constant, and this is an assumption that increasingly fails in higher dimensions. While the same assumptions underlie the partition-based estimators, the more uniform frequency, but variable macrostate size makes this assumption more realistic.

It is worth saying that the DSE estimator may give significantly different results with different choices of $K$, and potentially providing a better estimate with higher $K$. However, choosing $K$ by itself is challenging, especially in the absence of information about the underlying distribution or a known ground truth. In conclusion, only the KSG estimator shows poor performance in high-dimensional data, while all other estimators were able to capture the general trend, with notably superior performance from the $\pi$ estimator. 

\subsection{Synchronization}
A primary motivation for the development of these sample-based partitioning approaches to entropy estimation was the failure of previous MI estimators to incorporate the impact of outliers, especially in applications where outliers were known to be the most informative data points for certain types of analysis.  The study of the process of synchronization between coupled oscillators is a classic example of such an application because the (often short) transient period differs largely from the overall orbit of the synchronous state, which is confined to a small neighborhood of a low dimensional sub-manifold of the state space. 

In the study of synchronization between two identical dynamical systems, each system is defined on some region $\mathscr{D}\subset\mathbb{R}^d$ with additional coupling term(s) being added to the differential (or discrete) equations that define their motion.  Trajectories are then represented in the joint space $\mathscr{D}\times\mathscr{D}\subset\mathbb{R}^{2d}$, and once synchronized (excluding the transient period), the dynamics of the system in the joint space will essentially occur along the Kronecker product of the vector $\mathbb{1}=[1,1]^T$ with a single orbit in the state space $\mathscr{D}$.

Thus, the transient period (those iterations beginning from some non-synchronized initial conditions until the first time step where the two states are within some tolerance of the synchronization manifold) consists of what may be considered outliers in the joint space.  The portion of the recorded dynamics after synchronization occurs becomes the dominant contributor to traditional mutual information estimates. 

As such, when traditional MI estimators are used to study synchronization, the transient portion of the dynamics is largely ignored (in the case of histograms) or statistically de-emphasized (in the case of the KSG estimator).  Again, this has often been claimed as a benefit when you are only interested in the synchronized behavior, as the particular initial conditions are ignored in favor of a more consistent measure representing the synchronized portion of the dynamics.  This can, in fact, be desirable, for instance, when attempting to measure whether synchronization was achieved. But, when studying the process of synchronization or the effects of perturbations on an already synchronized system, all previous MI estimators are unable to provide any useful insight into the dynamic process of synchronization. 

In contrast, when using the partition-based MI estimators with the specific choice of measure of the form $g(L,F)=L/F$, the transient data (a set of outliers in the joint space) is now emphasized.  In fact, the emphasis on these potentially few outliers is weighted such that the partition-based MI estimators can differentiate between transient dynamics differing in even a single data point.  This sensitivity may be undesirable in other cases though, and so at the same time, the allowance for adaptation of these estimators through different choices for the form of measure enable a whole range of estimators that can effectively throttle the importance of outliers from extreme importance down to essentially ignored, just as with histogram estimates.  

The study of synchronization of chaotic oscillators is thus seen as an ideal context to explore the range of applications for this new set of MI estimators.  We begin with a simple unidirectional coupling on a pair of logistic maps, whose uncoupled dynamics is defined in terms of the parameterized function
\begin{equation}
\label{equ:logistic}
g_\alpha(x)=\alpha x(1-x),
\end{equation}
where $\alpha=4$ results in chaotic dynamics. That initial case, where only one dynamical system includes a coupling term, provides clear insight into the differences and effects of the choice in measure.  This is followed by the consideration of a symmetrically coupled pair of logistic maps, where the focus is more on the role of transient dynamics and an adaptive approach to ensure non-negative MI estimates. 

\subsubsection{Directed Coupling of Logistic Maps}
The ability to incorporate the impact of outliers in MI estimation and the resulting consequences is first explored in a simple directed (asymmetrical) coupling between a pair of logistic maps defined by
\begin{equation}
\label{equ:unicoupled}
\begin{aligned}
x_{t+1} &= g_4(x_t) \\
y_{t+1} &= (1-\sigma) g_4(y_t) + \sigma g_4(x_t),
\end{aligned}
\end{equation}
where $\sigma$ represents the coupling strength and the function $g_4(x)$ is defined in Equ~(\ref{equ:logistic}).  The mutual information between the two series $x_t$ and $y_t$, each consisting of an orbit of length $N=1000$, is estimated using the $I_\pi$ and $I_\square$ estimators as well as the histogram-based DSE and KSG estimators.  All four estimates are averaged over $1000$ random initial conditions in order to smooth out the effects of the different transient dynamics identified by the partition-based estimators, and one standard deviation is included as a shaded region around the mean in all figures.  

This averaging over various initial conditions is not generally needed for the traditional histogram and KSG estimators, especially when the dynamics synchronize, as their estimates largely do not depend on the initial conditions or the transient path to synchronization.  This observation in and of itself provides a perfect illustration of how the KSG estimator's being "robust to outliers" can be problematic for our interest in studying the path to synchrony.  

\begin{figure}[ht!]
    \centering
    \includegraphics[width=6.5cm]{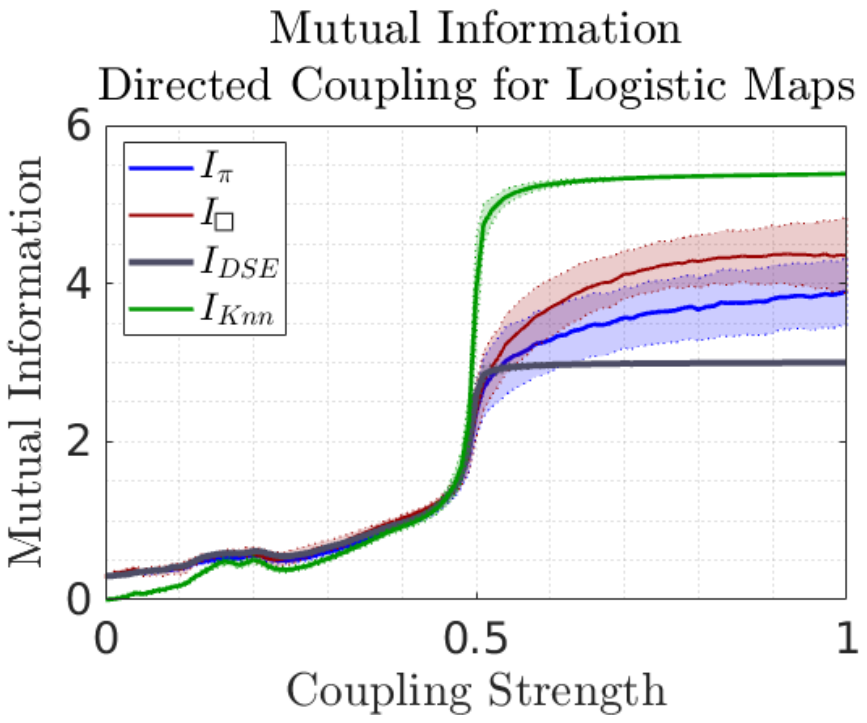}
    \caption{The mutual information is estimated by the box and $\pi$ estimators using $k=24$ along with the histogram-based DSE and the KSG approaches for the directed coupling of logistic maps of varying coupling strengths ranging from $y_{t+1}$ being entirely determined by $y_t$ to entirely determined by $x_t$.}
    \label{fig:Unicoupling}
\end{figure}

Despite the striking similarities in Fig.~\ref{fig:Unicoupling} between the two partition-based estimators and the more traditional estimates, there are several key differences.  Most notable is the variance of the partition-based estimators, which should not be a surprise and is explained by the inclusion of the transient dynamics in the estimate.  The other feature of note is the concavity of the region $\sigma\in(0.5, 1.0)$, where both the DSE and KSG estimates are essentially flat.  This may seem to indicate a problem with the new estimators, but upon deeper consideration, this feature is due entirely to the weighted incorporation of outliers into the estimation procedure.  Furthermore, the chosen measure of $\mu$, i.e. taking the form $g(L,F)=L/F$, means that information about the short transient period is emphasized regardless of a longer region of synchronized dynamics, since those much more highly populated states' measures are divided by the frequencies, reducing their importance.  This leads to what might be considered an over-emphasis on the impact of outliers, and in fact, the shorter the transient, the larger the mutual information estimate can be as the outliers increase in their perceived informativity.

Beginning with very small coupling strength, we find that both partition-based MI estimators largely coincide with the DSE estimate.  And while this value is not zero as the KSG estimator provides, this is not surprising given the potential for similar dynamics to result in low amounts of MI given a particular coarse-graining into macrostates.  The mutual independence test was developed to assist in determining whether these small values of MI are in fact anomalous~\cite{runge2018conditional}.  Regardless, we see the same general trend of slightly increasing MI as the coupling strength increases, including the notable bump for the same range of coupling strengths where both the DSE and KSG estimators increase due to near periodic dynamics of $y_t$.

More importantly, we see that once synchronization is reached, for $\sigma>0.5$, the differences in the transient dynamics prior to synchronization are registered by the partition-based approaches, whereas they are ignored by traditional estimators.  This feature, as indicated by the larger shaded regions defined by a standard deviation from the mean (over $1000$ runs), is consistent as the coupling strength increases.  While the transient dynamics get shorter with increased $\sigma$, in terms of the portion of the joint orbit away from the synchronization manifold, due to the growing relative importance placed on the contribution of the outliers, we see a general increase in MI until the point at $\sigma=1$, where after a single time step, $y_{t+1}$ is entirely defined by $x_t$.  Depending on the choice of parameter $K$, the box estimator will sometimes slightly reduce as $\sigma\rightarrow 1$, but this is simply a result of the inclusion of more empty state space in its entropy calculations.  The important point is that on any given set of initial conditions, the value of MI as compared to an orbit on the synchronized state, can provide insight into the process of that particular initial condition's path toward synchrony.

At this point, it is natural to ask whether the MI is equal to the entropy of the trajectory of $x_t$, since after one iteration, the two sequences are identical.  But again, due to the outsized weight placed on the single outlier in the joint space by the box and $\pi$ estimators, this will not be the case.  This serves as a reminder that these MI estimators do not treat each symbol with the same informational value.  Thus, here, the one observation that lies outside of the synchronization manifold contributes more to the overall measure of shared information in the symbolic representations of the dataset since the importance of each symbol is determined by $g(L,F) = L/F$.  

We wish to remind the reader that as with most information measures, it is the differences in values that are the most informative since the estimates rely on choices in the discretization process.  So, the MI estimate that includes the transient is best compared with the value of MI absent the transient period in order to extract meaningful measures of perturbation. 

\begin{figure}[ht!]
    \centering
    \includegraphics[width=6.5cm]{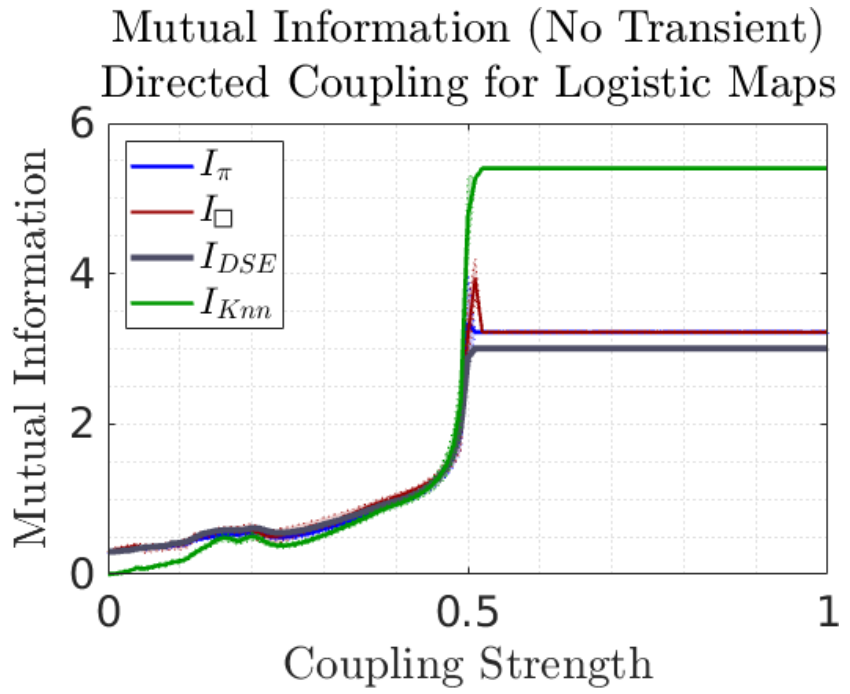}
    \caption{The MI estimates as shown in Fig.~\ref{fig:Unicoupling}, except the transient dynamics are skipped before recording the sequences for estimation of MI leading to the absence of variation in outcome in line with traditional estimators.}
    \label{fig:notransients}
\end{figure}

To further illustrate the dependence on transient dynamics, we plot the four MI estimates for the same dynamics again in Fig.~\ref{fig:notransients}, but skipping the transient dynamics by simulating the synchronization process for an orbit of length $N=2000$, followed by estimating MI only over the last $\Tilde{N}=1000$ points of the trajectory.  Here we see both the box and $\pi$ estimators cease to include the variation and essentially match the histogram DSE estimate even though the same measure $\mu$ of the form $g(L,F)=L/F$ was used as in Fig.~\ref{fig:Unicoupling}.  This is indicative of the fact that regardless of any differences in the basic entropy estimation, when considering information measures based on changes in entropy, those differences are largely negated~\cite{landau2013statistical}.

The additional spike at the cusp of synchronization for the box (and less so for the $\pi$) estimator is indicative of the inclusion of more outliers as the majority of the dynamics are beginning to synchronize.  Under the choice of measure $g(L,F)=L/F$, those joint points that are nearly, but not fully synchronized are treated as more informative, and since their symbolic representations are largely shared across marginal spaces, we find an notable increase in partition-based MI estimates.  

\begin{figure}[ht!]
    \centering
    \includegraphics[width=6.5cm]{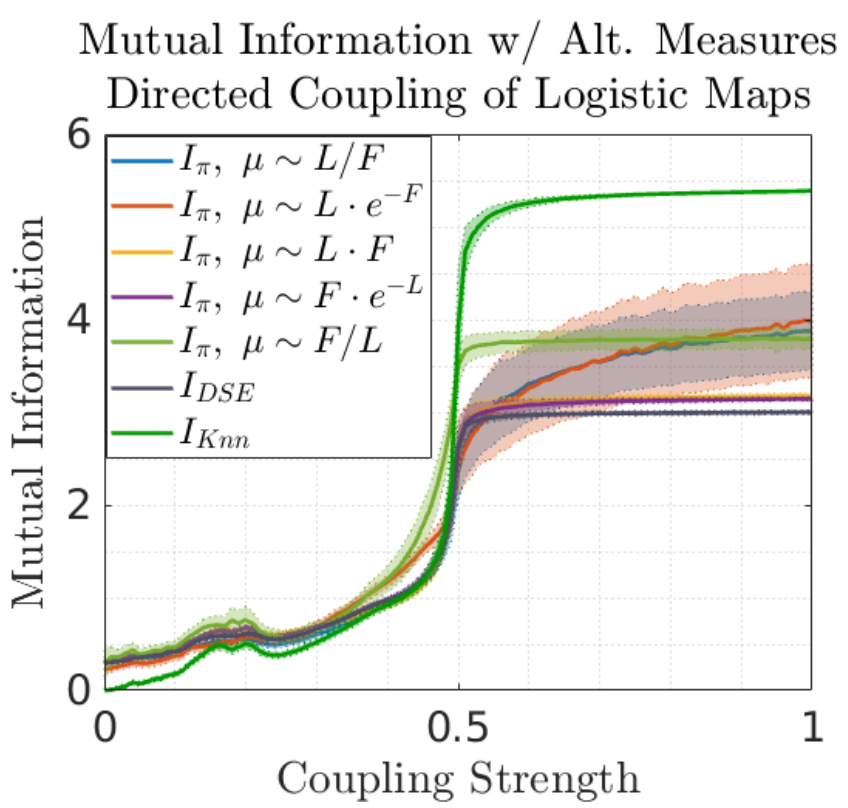}
    \caption{The MI estimates using the $\pi$ partitioning scheme with various choices in the form of the measure along with the DSE and KSG estimates for comparison.}
    \label{fig:AllPi}
\end{figure}

For the purpose of studying transient dynamics, the measure of the form $g(L,F)=L/F$ is most informative, however, the MI was estimated for the same data again using the $\pi$ estimator, but employing several alternative measures of different functional forms.  Fig.~\ref{fig:AllPi} shows plots of the MI as estimated by the $\pi$ estimator using a normalized measures of the forms: i) $g(L,F)=L/F$, ii) $g(L,F)=F\cdot e^{-L}$, iii) $g(L,F)=L\cdot F$, iv) $g(L,F)=L\cdot e^{-F}$, and v) $g(L,F)=L/F$ for the same coupling strengths for a direct comparison with the outcome from Fig.~\ref{fig:Unicoupling}.

With the $\pi$ estimator in particular, it is common to get a frequency vector, $F$, of the geometrically defined macrostates having large variation.  This can lead to an overemphasis on the outliers, but also presents a challenge in the presence of within tolerance repeated values, where the geometries associated with the repeated observations get reduced greatly.  We see here that the functional form of $g(L,F)=F\cdot e^{-L}$ can be useful in such cases, as was suggested previously.

Thus, for all subsequent figures, whenever an partition-based estimate of MI using the measure of the form $g(L,F)=L/F$ results in a negative value (which occurs most often in the presence of largely repeated values), we will automatically recompute the MI estimate using the form $g(L,F)=F\cdot e^{-L}$ to obtain a positive estimate that is still driven largely by the geometry, but more accurately incorporates the impact of variation in frequency.  While this emphasis on frequency with reduction of geometric influence by the exponential term leads to poor estimates for low coupling due to similar challenges with DSE, as will be seen in the next section, when periodic orbits are observed, the estimates of these two forms generally coincide for meaningfully large values of MI, making this adaptive choice in measure appropriate.

Next, perhaps as expected, we see that the choice of $g(L,F)=L \cdot e^{-F}$ is very similar to the specific volume measure, though it reduces the impact of variation in frequencies.  Alternatively, the measure with a form $g(L,F)=L\cdot F$ coincides to a large degree with the DSE estimate by effectively ignoring outliers due to the low frequencies that get associated with outlier regions, regardless of their geometric size. However, it does provide a more accurate estimate for low coupling due to the data-defined macrostates. Finally, we include a measure that can be thought of as "Lebesgue-like" local density in the form $g(L,F)=F/L$. Perhaps surprisingly, we find this to be a poor measure because it uses the frequencies of unequal bins to assign importance, but then counters the impact of the larger geometric regions by dividing by $L$.  It is not surprising then that this estimator results in a consistently unstable estimate where the variation is not dependent on the transient dynamics.  

Similar results to these hold for the box estimator as well, but in that case, the full domain $\mathscr{D}$ will always be included and the frequency vectors will be nonzero and generally have less variation (assuming few repeated values).  This can result in additional challenges in choosing the best measure for the analysis goal, especially if the data is sparse in regions of the marginal spaces, but as this is not the case here anyway, we chose to avoid confusion by its inclusion.

We now explore the interesting case of symmetrically coupled logistic maps where the emergent synchronous state is no longer defined by either of the initial conditions, $x_0$ or $y_0$.

\subsubsection{Symmetrically Coupled Logistic Maps}
\begin{figure}[ht!]
    \centering
    \includegraphics[width=8.1cm]{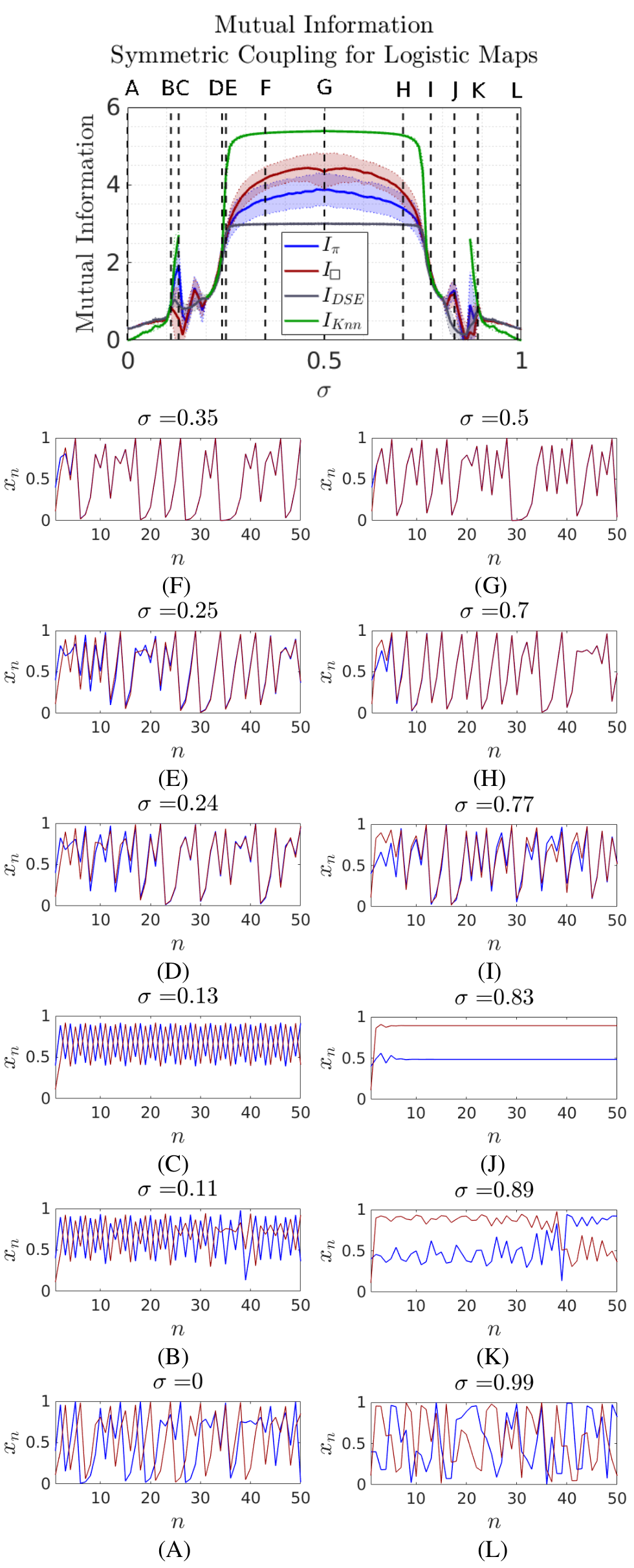}
    \caption{A comparison of the $\pi$ and box estimators with both the histogram DSE and the KSG estimates of MI for orbits of length $N=1000$ from coupled logistic maps (averaged over $1000$ random initial conditions); along with (A)-(L) the first $50$ iterations of the trajectory for the indicated coupling strengths starting from the initial condition $x_0=0.400$ and $y_0=0.111$.}
    \label{fig:BiLogistic}
\end{figure}
Consider a pair of symmetrically coupled chaotic logistic maps of the form:
\begin{equation}
\begin{aligned}
x_{t+1} &= (1-\sigma) g_4(x_t) + \sigma g_4(y_t)\\
y_{t+1} &= (1-\sigma) g_4(y_t) + \sigma g_4(x_t),
\end{aligned}
\end{equation}
where $\sigma$ represents the coupling strength and the function $g_4(x)$ is defined in Equ.~(\ref{equ:logistic}) as before.  This system provides a better view into analysis of the transients that can arise in coupled dynamics, since the synchronous state is not defined by either initial condition.  Varying the coupling strength through a range of values $\sigma\in[0,1]$, we again estimate the MI between the two resulting time series through different coupling regimes. 

An example trajectory of the initial $50$ iterations of each indicated coupling strength is plotted in Fig.~\ref{fig:BiLogistic} panels (A)-(L) with the initial conditions of $x_0=0.400$ and $y_0=0.111$ in each case.

In the absence of coupling (A), we again see two orbits of the logistic map for the pair of initial conditions, with very small values of $I_\square$ and $I_\pi$ as compared to the DSE estimate.  Under small coupling strength, we begin to see periodic behavior emerge as the coupling strength increases past $\sigma=0.11$ (B) into the range where the KSG estimator fails due to highly repeated values in periodic trajectories.  

Although the standard measure~(\ref{equ:measure}) is still computable for the periodic trajectories, they begin to result in negative MI estimates in this region due to the divisions of geometries by the frequencies of highly repeated values.  Thus, for these coupling strengths, the MI is automatically recalculated using a measure of the form $g(L,F)=F\cdot e^{-L}$.  In fact, under this altered choice of measure, we find both the box and $\pi$ estimators can correctly differentiate between those coupling strengths that result in true periodic conditions.  This is something that is unique to these estimators, since depending on the parameter $K$, the equal-partitioning of the histogram estimator can fail to differentiate small deviations from true periodic trajectories.

Full synchronization begins to occur around $\sigma\approx 0.26$, but interesting variation in the length of the transient occurs before this value (C and D), which only the box estimator is able to accurately differentiate..  Once $\sigma>0.26$, we find that both the histogram and KSG estimators plateau as they are unable to account for the outliers in the joint space that represent the transient dynamics during synchronization.  However, both partition based estimators are able to differentiate the increases in mutual information from the change of the transient of even a few iterations (between F and G), even when computed over the entire orbit of length $N=1000$.  Again, this ability is a direct result of the use of the measure $\mu$ enabling a large weight assignment to the increased importance of the few outliers from the transient in the joint space. 

As the coupling strength increases past a certain value (H), the coupling begins to overshoot the difference in dynamics between iterations, leading to spontaneous desynchronization in (I), which then leads to a region where the system converges to a fixed point in (J).  Pushing the coupling past this balanced point, we find the emergence of chaotic coupled dynamics again in (K) and (L), though these orbits are not simply the logistic map and no longer synchronize. 

This example illustrates the power of the partition-based MI estimators to provide estimates for all coupling strengths through adaptive measure choice and to provide insight into how a given initial condition synchronized; something that is unique to these estimators.

\section{Conclusion}
\label{sec:conclusion}
The basic concept of geometric partition entropy presented in~\cite{diggans2022geometric} has been transformed into a more mature framework for the study of entropy in high dimensional information theory in a way that enables the incorporation of informative outliers.  The Voronoi-based definition grounds the new theory in the historical context of Kolmogorov and Sinai, and the various approximation schemes enable computationally efficient and versatile estimators that can be applied in a wide range of contexts, including data-driven bounds on unbounded data.  The concept of GPE provides a more robust approach to data-driven entropy estimation in that it avoids bias to $\log_2{(N)}$, providing a more informative and meaningful value as the parameter $K\rightarrow N$. 
The subsequent mutual information estimators based on the concepts of GPE are shown to outperform traditional estimators in several respects, especially as applied in particular contexts.  The $\pi$ estimator is orders of magnitude more efficient in time complexity than even the histogram estimator, and is able to provide better estimates under sparse data, however it remains subject to the curse of dimensionality more generally. 

Most importantly, this new approach to mutual information estimation allows for the inclusion of informative outliers, and in fact is amenable to user-defined importance of outliers based on the chosen measure $\mu$. For instance, we have shown that certain choices of measure can result in the same de-emphasis of outlier impact as traditional estimates, and so these approaches provide a new dynamic set of tools for information theory, specifically for the study of complex systems, but also more broadly.

Through the incorporation of informative outliers in the analysis of dynamical systems, the partition-based mutual information estimators enable a paradigm shift in the study of synchronization dynamics by including a measure of how the transient dynamics contribute to the shared information.  The ability to study transient dynamics is applicable to analysis of disruptions to powergrids and many other important systems.  Of particular interest to the authors will be applications in weather prediction and other chaotic and switching dynamical systems.

\section{Acknowledgements}
We are grateful for support from the Air Force Office of Scientific Research under contract 23RICOR001.

\section{Data Availability}
 Data and relevant code for this research work are stored in GitHub: https://github.com/almomaa/GGPE4MI and have been archived within the Zenodo repository:  https://doi.org/10.5281/zenodo.13131645

\bibliographystyle{unsrt}
\bibliography{ggpe4mi}
\end{document}